\documentclass[aps,pre,twocolumn,groupedaddress]{revtex4-2}
\ifx\pdfoutput\undefined
\usepackage{graphicx}
\else
\usepackage{epstopdf}
\usepackage{epsfig}
\usepackage{amssymb,amsmath,stmaryrd,tabularx}
\usepackage{wrapfig}
\usepackage{bm}
\usepackage{ulem}
\usepackage{xcolor}
\fi

\usepackage[center]{subfigure}

\begin{document}

\title{A Field-Theoretic Model for Chemotaxis in Run and Tumble Particles}

\author{Purba Chatterjee and Nigel Goldenfeld}

\affiliation{Department of Physics, University of Illinois at Urbana-Champaign, Loomis Laboratory of Physics, 1110 West Green Street, Urbana, Illinois, $61801-3080$}

\date{\today}

\begin{abstract}
In this paper we develop a field-theoretic description for run and tumble chemotaxis, based on a density functional description of crystalline materials modified to capture orientational ordering. We show that this framework, with its in-built multi-particle interactions, soft-core repulsion and elasticity is ideal for describing continuum collective phases with particle resolution, but on diffusive timescales. We show that our model exhibits particle aggregation in an externally imposed constant attractant field, as is observed for phototactic or thermotactic agents. We also show that this model captures particle aggregation through self-chemotaxis, an important mechanism that aids quorum dependent cellular interactions.
\end{abstract}

\maketitle

\section{Introduction}\label{sec1}

Active matter consists of self-propelled particles which convert stored energy into directed motion, thus keeping
the system perpetually driven out of equilibrium. Interactions among such self-propelled particles (SPPs) give rise to novel collective behavior with no equilibrium analogue. A class of collective phenomenon that is widely studied is chemotaxis, the collective migration of active agents towards favorable environments \cite{Keller1970,Adler1966,Budrene1995,Maeda1976,Mittal2003}. This chemotactic motility depends on the ability of organisms to sense spatial gradients and reorient themselves in the direction of increasing attractant concentration, which  could be a chemical or an environmental stimulus such as light or heat. While chemotaxis is observed in a wide variety of organisms, its manifestation in bacteria is especially interesting because they are physically too small to sense spatial gradients directly. Certain bacteria like \textit{E. coli} overcome this handicap by adopting a ``run and tumble" approach, in which they move ballistically in a given direction for a while, then pause to tumble in space and choose a new direction forward. Directed motion is achieved by reducing tumbling rates when a favorable direction is identified. Chemotaxis as a means for signaling gives rise to many complex spatial patterns, the most basic and inevitable among which is aggregation \cite{Park2003,Park2003a,Budrene1991,Woodward1995,Budrene1995,Mittal2003,Adler1966,Saragosti2011}. 

\begin{figure}
\begin{center}
\includegraphics[scale=0.25]{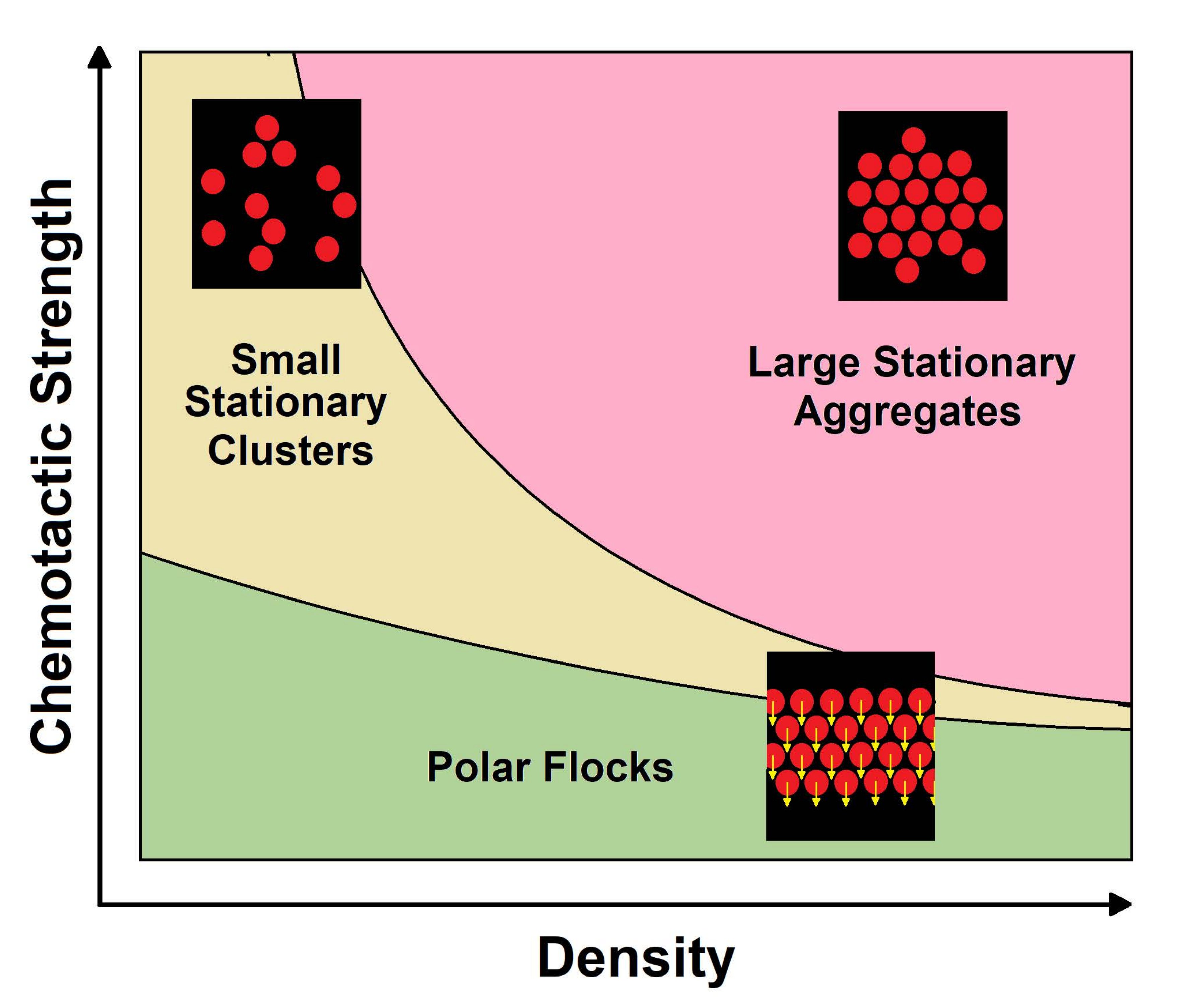}
\end{center}
\caption {\small{(Color online). Schematic phase-diagram for a system of run and tumble particles in the phase-space of density and chemotactic strength.}}
    \label{fig0}
\end{figure}

How do the interactions between individual chemotactic agents give rise to such complex collective phases? The classical Keller-Segel model that is most commonly used to study chemotaxis is a phenomenological continuum description of the interaction of the agents with their environment, but ignores microscopic interactions among the agents themselves. Furthermore, it is purely deterministic. As such it is not able to address questions related to the spatial structure and correlations between agents. A mesoscopic description that has recently been shown to successfully describe collective phenomenon in active systems is the active Vacancy Phase Field Crystal (VPFC) model \cite{Alaimo2016,Menzel2014}. This is a field-theoretic model in which the density field is driven by a free energy that imposes a locally periodic ground state. The peaks of the local density field are interpreted as `particles', which diffuse and collide with other particles on the emergent lattice. This allows us to study dynamic processes with particle resolution but on diffusive time scales, making it much more efficient than Molecular Dynamics (MD) simulations. This model has been shown to capture long-range orientational ordering in SPPs by successive alignment of particles through inelastic collisions \cite{Alaimo2016,Menzel2014}. Such interactions are inherently multi-particle since each particle interacts with all its neighbors on the lattice. This has been recently shown to be important for observing flocking in active systems \cite{Suzuki2015,Chatterjee2019}. When modified to describe chemotaxis, we expect the active VPFC framework to generate particle aggregation through the collective searching of attractant gradients. The higher the density or the strength of chemotactic interactions, the greater the chance of forming large stationary aggregates, due to the cooperative aligning interactions among closely spaced neighbors. For low chemotactic strengths we expect to recover the dynamics of the original VPFC model, i.e. flocks with long-range orientational order. For high chemotactic strength and low densities, we expect the formation of scattered clusters but no large aggregates. In the phase-space of density and chemotactic strength, this can be represented by the schematic phase diagram shown in Fig.~\ref{fig0}.

It is important to note here that although in this paper we focus chiefly on run and tumble particles, the mechanism for aggregation prescribed by our model can be generalized to a large class of social organisms. Insect swarms, for example, have been shown to form stable cohesive groups even in the presence of environmental noise \cite{Kelley2013,Puckett2014,Puckett2014a}, just like chemotactic bacteria. Aggregation in social organisms is generally attributed to individual agents interacting via long-range, centrally attractive, gravity-like forces, making them analogous to self-gravitating Brownian particles \cite{Okubo1986,Gorbonos2016,Chavanis2007}. Recently it was shown that such long-range forces can emerge from the interplay between two modes of flight in midges- the first a low frequency maneuver consisting of mainly to and fro movement (flying), and the second a high frequency nearly-harmonic oscillation in synchrony with another midge (hovering) \cite{Puckett2015,Reynolds2019}. There is then a clear analogy between these midges and run and tumble bacteria. 

Our broader goal thus is to establish the active VPFC model as a framework for studying collective phases in a wide variety of active systems. The usefulness of this framework lies in its ability to capture the emergence of macroscopic collective behavior through multi-body interactions among individual agents. Most important is the contribution it can make in studying the emergent material-properties of these collective phases, such as elasticity and surface tension. Insect swarms  have been shown to exhibit macroscopic mechanical properties, such as collective viscoelasticity as a response to oscillatory visual stimuli, as well as solid-like properties such as a finite Young’s modulus and yield strength \cite{Ni2016,Vaart2019,Reynolds2019a}. Schools of phototactic freshwater fish exhibit linear elasticity when spatially and temporally varying light fields are used as stressors \cite{Pokhrel2019}. In general the cohesivity of individuals within a stable aggregate presupposes emergent rigidity. An aggregate of particles may be thought of as achievement of local crystalline order. Here, we argue, lies the advantage of our field-theoretic framework. The starting point of the active VPFC model is the equilibrium Phase Field Crystal (PFC) model \cite{Elder2002}, which is a minimal description for crystalline materials. It is the PFC free energy functional that imposes crystalline order on the evolving density field by penalizing deviations from spatial periodicity. This not only makes it possible for us to interpret the peaks of the density field as `particles', and study their dynamics, but the resulting density field also retains the crystallographic and elastic properties of the material in question. This is in contrast to phase field descriptions such as the Keller-Segel model, which do not permit the study of either individual particle dynamics, or emergent mechanical properties. The equilibrium PFC model is implicitly elastic, because any deformation of the underlying crystalline lattice increases the free energy \cite{Chan2009,Huter2016}. Small deformations result in a quadratic increase in energy, thus capturing linear elasticty. If the deformation is large, the elastic response is non-linear. This underlying elasticity of the equilibrium theory is inherited by the non-equilibrium active VPFC framework, which extends the equilibrium PFC model to describe `active' crystals \cite{Alaimo2016,Menzel2014} where individual components are self-propelled and penalized or rewarded for aligning with their neighbors. This is then the natural way to combine the study of emergent collective behavior in active systems with the investigation of the material properties associated with a particular collective phase. For example, the emergent elasticity of chemotactic aggregates can be studied with the active VPFC model, by introducing deformations on the locally crystalline steady state. 

As a demonstration of the efficiency of this approach, here we develop an active VPFC model for run and tumble chemotaxis, which retains the microscopic details of interactions among individuals while also being numerically tractable. The paper is organized as follows. In section~\ref{sec2} we derive the active VPFC model from a density functional description of individual particles, in order to develop the general framework on which modifications can be made to study special cases of collective behavior in active systems. We show that this model is able to describe long range orientational ordering in active systems, through inelastic collisions among individual agents. In section~\ref{sec3} we derive the active VPFC chemotaxis model and study it for two different cases of attractant fields, one constant and externally imposed, and the other secreted by the particles themselves and diffusing in the same medium. We show that in both these cases our model captures the essential features of particle aggregation that is expected. In section~\ref{sec4} we end with a discussion on further applications of the active VPFC framework, especially to study emergent material properties of collective phases. 


\section{The active Vacancy Phase Field Crystal (active VPFC) model}\label{sec2}

We will first derive the active VPFC model starting from density functional theory. The time evolution of the single particle density $f(\hat{\mathbf{u}},\mathbf{r},\tau)$ of agents at position $\mathbf{r}$ and time $\tau$, moving along the direction of the unit vector $\hat{\mathbf{u}}$, is given in $2$D by:
\begin{equation}\label{eq2.1}
\partial_{\tau} f(\hat{\mathbf{u}})=D_T \nabla^2 \Bigg(\frac{\delta \mathcal{F}}{\delta f(\hat{\mathbf{u}})}\Bigg) +\tilde{D}_R\nabla_{\hat{\mathbf{u}}}^2 \Big(\frac{\delta \mathcal{F}}{\delta f(\hat{\mathbf{u}})}\Big) -\tilde{v}\boldsymbol{\nabla}.\big(f(\hat{\mathbf{u}})\hat{\mathbf{u}}\big),
\end{equation}
where $D_T$ and $\tilde{D}_R$ are the translational and rotational diffusion coefficients respectively. The equilibrium free energy functional $\mathcal{F}$ that drives the single particle density will be specified later. Self-propulsion is introduced by the last advective term in Eq.~\ref{eq2.1}, and the constant velocity $\tilde{v}$ is a measure of the activity of the system.

The first and second moments of the single particle density $f(\hat{\mathbf{u}},\mathbf{r},\tau)$ are the local density field $\tilde{\rho}(\mathbf{r},\tau)$ and the local polarization field $\mathbf{\tilde{P}}(\mathbf{r},\tau)$:
\begin{align}
\tilde{\rho}(\mathbf{r},\tau)&=\frac{1}{2\pi}\int d\hat{\mathbf{u}} \hspace{5pt} f(\hat{\mathbf{u}}),\label{eq2.2}\\
\mathbf{\tilde{P}}(\mathbf{r},\tau)&=\frac{1}{\pi}\int d\hat{\mathbf{u}} \hspace{5pt}\hat{\mathbf{u}} f(\hat{\mathbf{u}})\label{eq2.3}.
\end{align}
Expanding the single particle density in terms of its moments upto the second order, we have
\begin{equation}\label{eq2.4}
f(\hat{\mathbf{u}},\mathbf{r},\tau)=\tilde{\rho}(\mathbf{r},\tau) + \mathbf{\tilde{P}(\mathbf{r},\tau)}.\hat{\mathbf{u}}.
\end{equation}
Inserting this expansion into Eq.~\ref{eq2.1}, we can derive dynamical equations for the moments $\tilde{\rho}$ and $\mathbf{\tilde{P}}$, 
\begin{align}
\partial_{\tau} \tilde{\rho}&=D_T \nabla^2\Big(\frac{\delta \mathcal{F}}{\delta \tilde{\rho}}\Big) -\frac{\tilde{v}}{2}\boldsymbol{\nabla}.\mathbf{\tilde{P}},\label{eq2.5}\\
\partial_{\tau} \mathbf{\tilde{P}}&=D_T\nabla^2 \Big(\frac{\delta \mathcal{F}}{\delta \mathbf{\tilde{P}}}\Big) -\tilde{D}_R\Big(\frac{\delta \mathcal{F}}{\delta \mathbf{\tilde{P}}}\Big) -\tilde{v}\boldsymbol{\nabla}\tilde{\rho}.\label{eq2.6}
\end{align}
The free energy functional $\mathcal{F}$ is specified in terms of the two order parameters $\tilde{\rho}$ and $\mathbf{\tilde{P}}$, 
\begin{equation}\label{eq2.7}
\mathcal{F}[\tilde{\rho},\mathbf{\tilde{P}}]=\mathcal{F}_{VPFC}[\tilde{\rho}]+\mathcal{F}_{\mathbf{\tilde{P}}}[\mathbf{\tilde{P}}].
\end{equation}
The first contribution to $\mathcal{F}$ is $\mathcal{F}_{VPFC}[\tilde{\rho}]$, a functional of the order parameter $\tilde{\rho}$ only. It is given by
\begin{align}\label{eq2.8}
\mathcal{F}_{VPFC}[\tilde{\rho}]&=\int d\mathbf{r}\hspace{5pt}\Bigg[\frac{\tilde{\rho}}{2}\bigg[-\epsilon + (q_0^2+\nabla^2)^2\bigg]\tilde{\rho} +\frac{\lambda\tilde{\rho}^4}{4}\Bigg]\nonumber \\
&+ \frac{1}{6}\int d\mathbf{r}\hspace{5pt} \tilde{H} (|\tilde{\rho}|^3-\tilde{\rho}^3).
\end{align}
The first term of $\mathcal{F}_{VPFC}$ corresponds to the equilibrium PFC model, which describes dynamic processes in crystalline materials \cite{Elder2002}. The PFC model has been successfully implemented to realistically describe a wide range of material phenomenon such as elasticity, epitaxial growth, and defect, crack and fracture dynamics. This term of the free energy $\mathcal{F}_{VPFC}$ is different from the standard Landau-Ginzberg free energy, because it penalizes deviations from periodicity and not deviations from uniformity. The result is that there is a perfectly crystalline ground state about which distortions of various types can occur with an energy cost. It is this equilibrium PFC part of the total free energy functional $\mathcal{F}$, that encodes the elastic response of the system to mechanical strain, as will be shown in section~\ref{sec4}. Here $q_0$ is the inverse lattice spacing and $\epsilon$ is a control parameter which corresponds to a dimensionless temperature. In $2$D, this part of the free energy is minimized by striped, hexagonal or constant density profiles, depending on the temperature $\epsilon$ and the average density $\bar{\rho}$. The second term of $\mathcal{F}_{VPFC}$ penalizes negative values for $\tilde{\rho}$, $\tilde{H}$ being a large positive penalty. This allows for the description of individual particles with $\tilde{\rho}$ acting as a physical density \cite{Chan2009a,Alaimo2016}. 

The second contribution to $\mathcal{F}$ is $\mathcal{F}_{\mathbf{\tilde{P}}}[\mathbf{\tilde{P}}]$, a functional of the local polarization $\mathbf{\tilde{P}}$ only. It is given by,
\begin{equation}\label{eq2.9}
\mathcal{F}_{\mathbf{\tilde{P}}}=\int d\mathbf{r} \hspace{5pt} \Big[C_1 |\mathbf{\tilde{P}}|^2 + \tilde{C}_2(|\mathbf{\tilde{P}}|^2)^2\Big].
\end{equation}
This free energy functional promotes or penalizes alignment of self-propulsion directions depending on the sign of $C_1$. This incorporates into the active VPFC model, elements of continuum hydrodynamic descriptions of active matter like the Toner-Tu model \cite{Toner1995}, allowing us to observe emergent orientational order \cite{Menzel2014}. 

The chemical potential gradient associated with $\mathcal{F}_{VPFC}$ drives the time evolution of the local density, and that associated with $\mathcal{F}_{\mathbf{\tilde{P}}}$ drives the time evolution of the local polarization direction. Here we note that the free energy $\mathcal{F}[\tilde{\rho},\mathbf{\tilde{P}}]$ can be generalized to include correlations between $\tilde{\rho}$ and $\mathbf{\tilde{P}}$. However, as we will show, this minimal form of the free energy is nevertheless able to capture the collective dynamics of self-propelled agents.

Inserting the form of $\mathcal{F}[\tilde{\rho},\mathbf{\tilde{P}}]$ into Eqs.~\ref{eq2.5} and \ref{eq2.6}, and applying the following rescaling rules,
\begin{align}
\tau&=t/D_T ,\hspace{8pt}\tilde{v}=\sqrt{2}D_Tv/M_0 ,\hspace{8pt}\tilde{\rho}=M_0\rho,\hspace{8pt}\lambda=1/M_0^2,\nonumber\\ 
\tilde{H}&=H/M_0^2,\hspace{8pt}\mathbf{\tilde{P}}=\sqrt{2}\mathbf{P},\hspace{8pt}\tilde{C}_2=C_2/2,\hspace{8pt}\tilde{D}_R=D_TD_R,\label{eq2.10}
\end{align}
we get the final form of our active VPFC model,
\begin{align}
\partial_t \rho &= M_0\nabla^2 \Bigg[\Big[-\epsilon+\big[q_0^2+\nabla^2\big]^2\Big]\rho +\rho^3 + H\rho\big[|\rho|-\rho\big]\Bigg]\nonumber\\
&- v\boldsymbol{\nabla}.\big(\rho\mathbf{P}\big),\label{eq2.11}\\
\partial_t \mathbf{P}&=\nabla^2 \Big[C_1\mathbf{P}+C_2\mathbf{P}^3\Big] - D_r\Big[C_1\mathbf{P}+C_2\mathbf{P}^3\Big] - v\boldsymbol{\nabla}\rho.\label{eq2.12}
\end{align}
Here $M_0$ is the particle's mobility. In general the mobility of an agent can vary with local population density, and this property has been shown to generate spatially patterned collective phases in many biological systems, including but not limited to mussel beds and ant-colonies \cite{Liu2013,Theraulaz2002,Liu2016}. For the sake of simplicity, here we assume that the mobility $M_0$ is independent of the density. However, in section~\ref{sec3}$B$, we will show that in the case of run and tumble particles that perform self-chemotaxis, such density dependent mobility emerges from the mutual interactions of particles even with a constant value for $M_0$. Small values of $M_0$ result in slightly elliptical shapes for particles, but a relatively large value of $M_0$ restores their circular shape \cite{Alaimo2016}. In Eq.~\ref{eq2.11} we modified the advection term for $\rho$ to allow the the particle's advection speed to depend on the local density. This more classical form for advection stabilizes individual particles and allows us to better observe the dynamics at extremely low densities \cite{Alaimo2016}.

\begin{figure*}
\begin{center}
\includegraphics[scale=0.125]{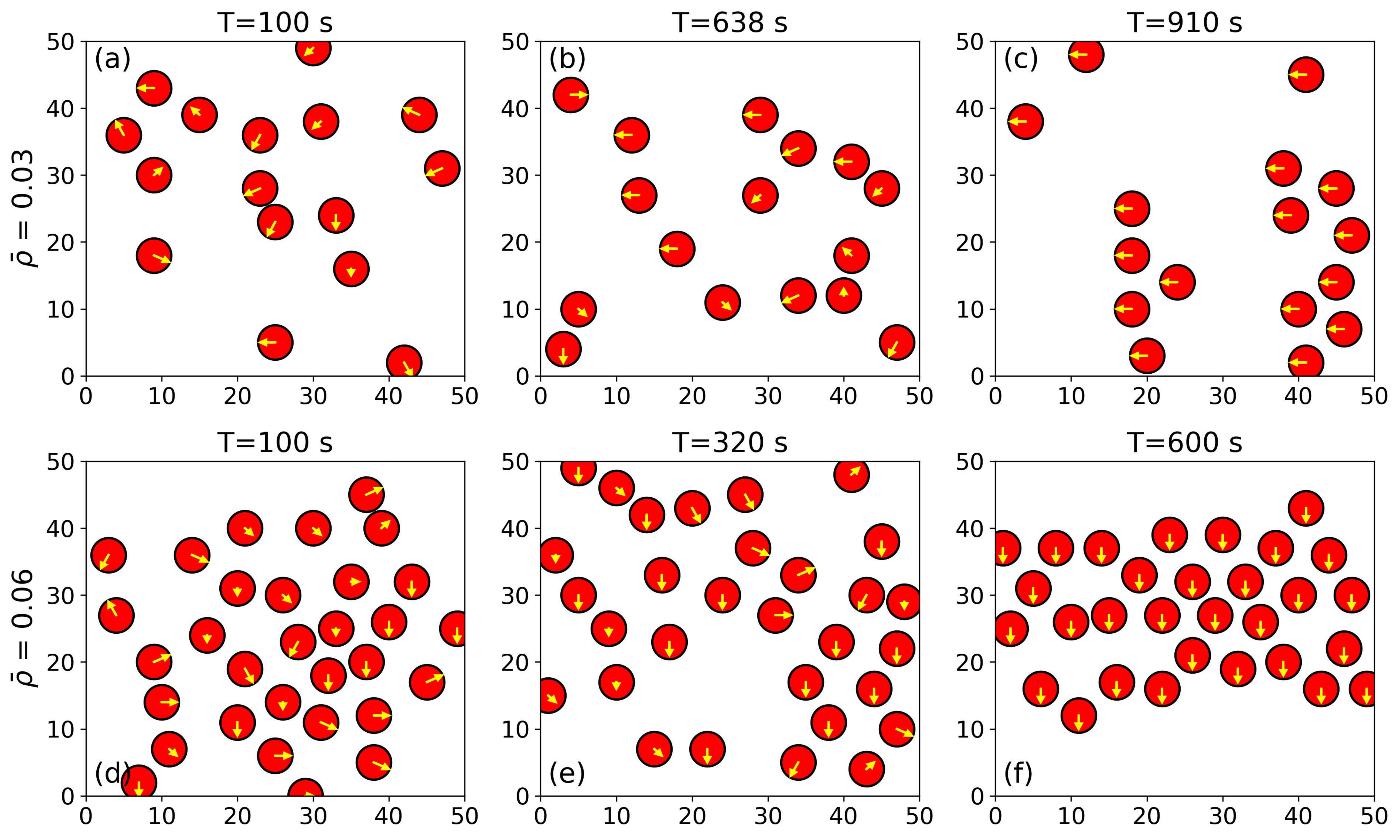}
\end{center}
\caption {\small{(Color online). Snapshots of the density profile at three different times, for average densities $\bar{\rho}=0.03$ (a, b, c), and $\bar{\rho}=0.06$ (d, e, f). Initially, particles move in random directions, but as more collisions take place over time, domains of particles travelling in the same direction appear, and eventually through collisions among domains all the particles travel in a single spontaneously chosen direction. $L=50,M_0=5,\epsilon=0.9,q_0=1,H=1500,v=2,C_1=0.2,C_2=0,D_R=1$.}}
    \label{fig1}
\end{figure*}

A version of this active VPFC model has been shown to successfully describe the dynamics of individual SPPs as well as emergent collective phenomenon in systems of SPPs such as flocking, vorticity, and boundary dependent oscillatory motion \cite{Alaimo2016}. Here we simulate the active VPFC model (Eqs.~\ref{eq2.11} and ~\ref{eq2.12}) considering a square domain of size $L\times L$, and periodic boundary conditions. Fig.~\ref{fig1} shows simulation results for two different average densities $\bar{\rho}$. The top panel (a, b, c) shows snapshots of the density profile at three different times for a low density system, while the bottom panel (d, e, f) shows the same for a high density system. In each of these snapshots, individual particles are represented by red solid circles and correspond to peaks of the local density field. The velocity of each particle is represented by an arrow whose length and direction is a measure of the speed and the average polarization of the particle respectively. In both the low as well as the high density cases, we see that initially there is no order and the particles move randomly in all directions (Fig.~\ref{fig1} (a, d)). After a while, particles start colliding with each other and form domains travelling in a particular direction (Fig.~\ref{fig1} (b, e)). Domain formation is mediated by the inelastic collisions between soft-core particles. When two particles collide, the transverse components of their velocities are depleted, because they cannot get closer to each other than their diameter (given by the lattice spacing). This assists in aligning the two particles, which then collide with other particles to form domains. When two such travelling domains collide, their constituent particles change their orientation till all of them are travelling in the same direction. Eventually, coarsening of domains results in all particles in the arena being orientationally ordered and travelling collectively in the same spontaneously chosen direction (Fig.~\ref{fig1} (c, f)). The higher the density, the greater the frequency of collisions, and the shorter the time taken to achieve complete orientational order. It is also interesting that given sufficient time, the particles arrange themselves in a travelling band, a feature very commonly observed in active systems that flock \cite{Schaller2010,Solon2013}. It is thus clear that the active VPFC model is able to capture the emergence of long range orientational order through interactions among active agents. 


\section{Active VFPC chemotaxis}\label{sec3}

Having demonstrated the efficiency of the active VPFC model in describing dynamical behavior in systems of SPPs, and the various advantages it offers, we propose to modify it to study ``run and tumble" chemotaxis in bacteria. In this, bacteria search for (are repelled by) a chemical attractant (repellant) diffusing in the same medium, by switching between running in approximately straight paths for a duration $\tau_{run}$, and tumbling to change direction in between runs at a rate $\alpha=\tau_{run}^{-1}$. The prototypical example is the bacterium \textit{Escherichia coli}, which is a tiny flagellated organism about $2 \mu m$ in length and $1 \mu m$ in diameter. During runs, its flagella rotate counterclockwise in a synchronous manner to propel it forward at a speed of $\approx 20 \mu m/s$, for an average duration of about $1 s$ \cite{Schnitzer1993}. In contrast, after every run, the bacterium tumbles for an average duration of about $0.1 s$, during which its flagella rotates asynchrously in a clockwise sense, keeping the bacterium held in place while it finds a new direction to move forward in. Alternating between runs and tumbles results in a random walk, where tumbles are analogous to collisions in molecular diffusion. However it is a biased random walk because when it is moving up (down) the gradient of attractant (repellent) concentration, a bacterium reduces the rate of tumbling $\alpha$ to allow for longer runs in that direction. Moreover, the new direction decided upon after tumbling is more likely to be in the forward hemisphere, a feature called directional persistence. In addition, the path taken during runs is not completely straight, but subject to rotational diffusion. 

To derive the active VPFC chemotaxis model for run and tumble particles, once again we start from density functional theory for the time evolution of the single particle density $f(\hat{\mathbf{u}},\mathbf{r},\tau)$ of agents moving along the direction of the unit vector $\hat{\mathbf{u}}$, in $2$D:

\begin{align}\label{eq3.1}
\partial_{\tau} f(\hat{\mathbf{u}})&=D_T \Delta \Bigg(\frac{\delta \mathcal{F}}{\delta f(\hat{\mathbf{u}})}\Bigg) +\tilde{D}_r\partial_{\hat{\mathbf{u}}}^2 \Big(\frac{\delta \mathcal{F}}{\delta f(\hat{\mathbf{u}})}\Big) -\tilde{v}\boldsymbol{\nabla}.\big(f(\hat{\mathbf{u}})\hat{\mathbf{u}}\big)\nonumber\\
&- \alpha(\hat{\mathbf{u}})f(\hat{\mathbf{u}}) + \oint d\hat{\mathbf{u}}^{'}\hspace{5pt} \alpha(\hat{\mathbf{u}}^{'})\gamma(\hat{\mathbf{u}},\hat{\mathbf{u}}^{'})f(\hat{\mathbf{u}}^{'}),
\end{align}
where $D_T$ and $\tilde{D}_r$ are the translational and rotational diffusion coefficients respectively, $\tilde{v}$ is the self-propulsion velocity and $\mathcal{F}=\mathcal{F}_{VPFC}[\tilde{\rho}]+\mathcal{F}_{\mathbf{\tilde{P}}}[\mathbf{\tilde{P}}]$ is the active VPFC free energy functional given by Eq.~\ref{eq2.7}, Eq.~\ref{eq2.8} and Eq.~\ref{eq2.9}. Specific to run and tumble particles, $\alpha(\hat{\mathbf{u}})$ is the rate at which an agent running along $\hat{\mathbf{u}}$ tumbles to find a new direction, and has the form
\begin{equation}\label{eq3.2}
\alpha(\hat{\mathbf{u}})=\tilde{\alpha}_0 +\boldsymbol{\tilde{\alpha}_1}.\hat{\mathbf{u}}=\tilde{\alpha}_0 -\chi(c,\tilde{v})(\boldsymbol{\nabla}c).\hat{\mathbf{u}}.
\end{equation}
Here $c$ is the concentration of the chemical and $\chi(c,\tilde{v})$ is the chemotactic sensitivity, which is positive for an attractant and negative for a  repellant. Thus, the particle reduces (increases) its tumbling rate when it is moving up the gradient of the attractant (repellant). For our model, we use the simplest form for $\chi(c,\tilde{v})$, which assumes that the chemotactic sensitivity is independent of local chemical concentrations, and directly proportional to the self-propulsion speed,
\begin{equation}
\chi(c,\tilde{v})=g\tilde{v}.\label{eq3.3}
\end{equation}
Angular correlations are accounted for by $\gamma(\hat{\mathbf{u}},\hat{\mathbf{u}}^{'})$, which is the probability that a particle initially running along $\hat{\mathbf{u}}'$ tumbles and chooses $\hat{\mathbf{u}}$ as the new direction. The angular correlation function satisfies 
\begin{equation}\label{eq3.4}
\int d\hat{\mathbf{u}}\hspace{5pt} \gamma(\hat{\mathbf{u}},\hat{\mathbf{u}}')=\int d\hat{\mathbf{u}}'\hspace{5pt} \gamma(\hat{\mathbf{u}},\hat{\mathbf{u}}')=1.
\end{equation}
\begin{equation}\label{eq3.5}
\int d\hat{\mathbf{u}}\hspace{5pt} \gamma(\hat{\mathbf{u}},\hat{\mathbf{u}}')\hat{\mathbf{u}}=\Gamma \hat{\mathbf{u}}'.
\end{equation}
Eq.~\ref{eq3.4} is a normalization condition that ensures that the integral over all possible outgoing directions is unity. Eq.~\ref{eq3.5} states that the outgoing direction is at an angle $\Gamma$ to the incoming direction on average,
\begin{equation}\label{eq3.6}
\langle \hat{\mathbf{u}}\rangle.\hat{\mathbf{u}}'=\Gamma\hat{\mathbf{u}}'.\hat{\mathbf{u}}'=\Gamma.
\end{equation}
Thus $\Gamma$ is a measure of the directional persistence of agents and is taken to be the mean cosine of the angle between successive directions of motion. The distribution of angular differences after a tumbling event is peaked at about $62^{\circ}$ for \textit{E.coli} \cite{Schnitzer1993}. Accordingly, we take $\Gamma=0.5$, corresponding to a mean difference between succesive angles of $\approx 60^{\circ}$. 

Once again the order parameters are the density field $\tilde{\rho}(\mathbf{r},t)$ and the polarization field $\mathbf{\tilde{P}}(\mathbf{r},t)$, the first and second moments of the single particle density $f(\hat{\mathbf{u}},\mathbf{r},t)$ respectively (Eqs.~\ref{eq2.2},~\ref{eq2.3}). Proceeding as we did in the last section, we expand the single particle density as $f(\hat{\mathbf{u}})=\tilde{\rho} + \mathbf{\tilde{P}}.\hat{\mathbf{u}}$. Inserting this expansion into Eq.~\ref{eq3.1} and applying the following rescaling rules,
\begin{align}
\tau&=t/D_T ,\hspace{8pt}\tilde{v}=\sqrt{2}D_Tv/M_0 ,\hspace{8pt}\tilde{\rho}=M_0\rho,\hspace{8pt}\lambda=1/M_0^2,\nonumber\\ 
\tilde{H}&=H/M_0^2,\hspace{8pt}\mathbf{\tilde{P}}=\sqrt{2}\mathbf{P},\hspace{8pt}\tilde{C}_2=C_2/2,\hspace{8pt}\tilde{D}_R=D_TD_R,\nonumber\\
\tilde{\alpha}_0&=D_T\alpha_0,\hspace{8pt}\boldsymbol{\tilde{\alpha}_1}=\sqrt{2}D_T\boldsymbol{\alpha_1}/M_0,\label{eq3.7}
\end{align}
we finally obtain the dynamical equations for an active VPFC chemotaxis model for run and tumble particles,
\begin{align}
\partial_t \rho&=M_0\nabla^2 \Bigg[\Big[-\epsilon+\big[q_0^2+\nabla^2\big]^2\Big]\rho +\rho^3 + H\rho\big[|\rho|-\rho\big]\Bigg]\nonumber\\
& -v\boldsymbol{\nabla}.\mathbf{P},\label{eq3.8}\\
\partial_t \mathbf{P}&=\nabla^2 \Big[C_1\mathbf{P}+C_2\mathbf{P}^3\Big] - D_r\Big[C_1\mathbf{P}+C_2\mathbf{P}^3\Big] \nonumber\\
&-v\boldsymbol{\nabla}\rho - \alpha_0(1-\Gamma)\mathbf{P}-\rho(1-\Gamma)\boldsymbol{\alpha_1}.\label{eq3.9}
\end{align}

In the following subsections we describe two different scenarios of positive chemotaxis that can be captured by this model. The first case corresponds to run and tumble particles in an attractant concentration field that remains constant in time and space, and is not degraded by diffusion or consumption. This scenario pertains to particles that prefer to reside in regions with high (or low) values of a particular environmental stimulus, such as light or heat. Such phototaxis or thermotaxis is in fact observed in a wide variety of biological systems, including but not limited to slime molds \cite{Bonner1950,Fisher1981}, bacteria \cite{Maeda1976,Ragatz1965}, mammalian spermatozoa \cite{Bahat2003}, nematodes \cite{Hedgecock1975} and larval fish \cite{Wurtsbaugh1988,Brockerhoff1995}. The second case looks at self-chemotaxis, where particles are attracted to a chemical that they themselves secrete, and which both diffuses and degrades depending upon its local concentration. Self-chemotaxis is observed in many biological systems, where individual particles act as moving sources of the attractant, and interact with one another to give rise to collective phases such as stable stationary aggregates, traveling bands and complex spatial patterns \cite{Park2003,Park2003a,Budrene1991,Woodward1995,Budrene1995,Mittal2003,Adler1966,Saragosti2011}. In fact the classical Keller-Segel model was first formulated to explain the aggregative properties of slime mold amoeba, mediated by self-chemotaxis towards \textit{aracsin} \cite{Keller1970}. 

We allow the active VPFC particles to stabilize for $\Delta T=50s$ before switching on the attractant field in the first case, and starting the production of the attractant in the second case. All simulations have been performed considering a square domain of size $L\times L$, and periodic boundary conditions. The simulation arena is initialized with random density and polarization profiles which have averages $\bar{\rho}$ and zero respectively.

\subsection{Constant Attractant Field}\label{subsec3.1}

\begin{figure*}
\begin{center}
\includegraphics[scale=0.2]{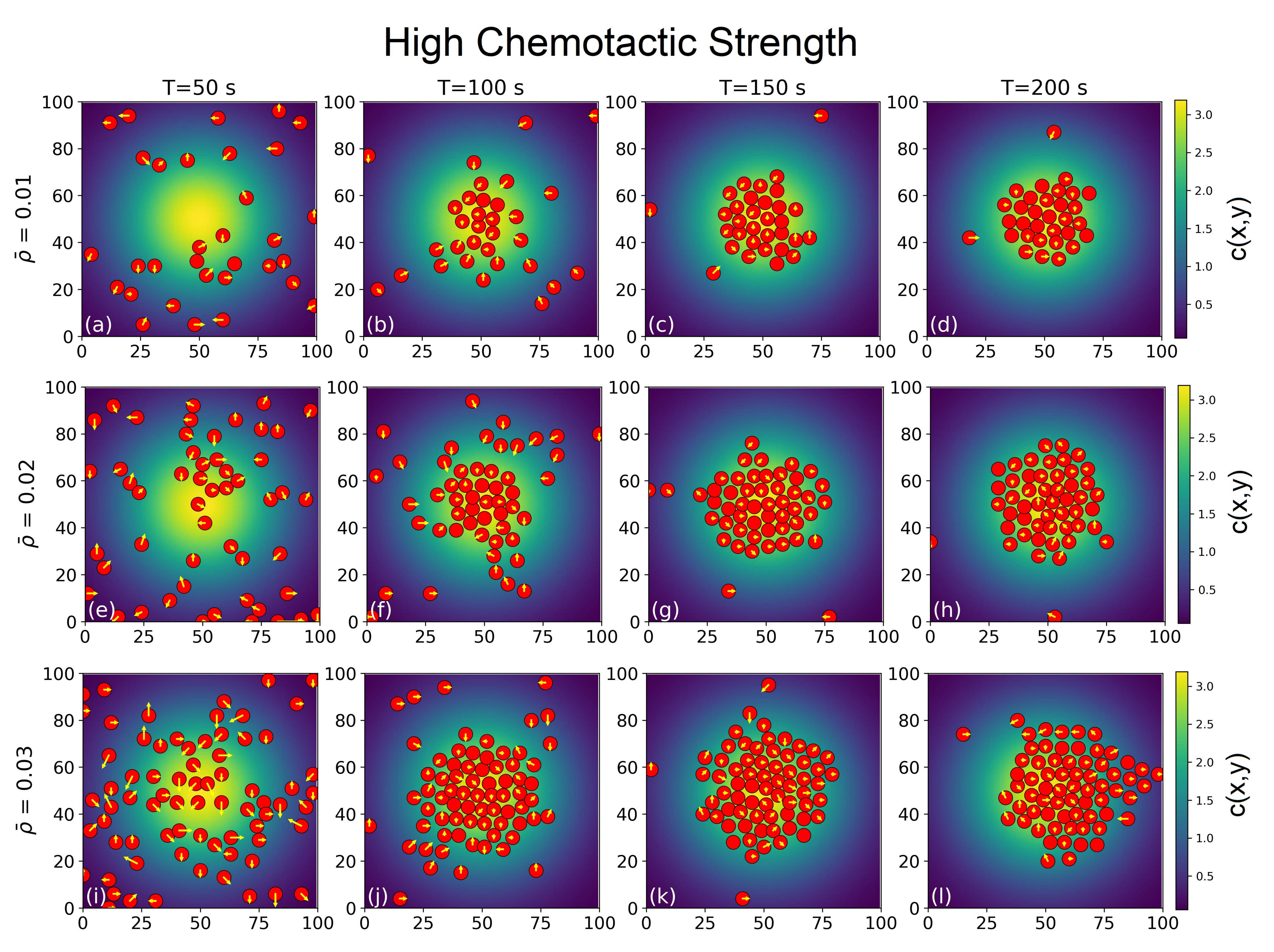}
\end{center}
\caption {\small{(Color online). Snapshots of density and velocity profiles at four different times $(50s, 100s, 150s, 200s)$, for average densities $\bar{\rho}=0.01$ (a, b, c, d), $\bar{\rho}=0.02$ (e, f, g, h), and $\bar{\rho}=0.03$ (i, j, k, l). The chemical concentration field $c(x,y)$ is represented by the colormap. Initially, particles move in random directions, but eventually aggregate in the center of the arena where the attractant concentration is maximum. Higher the density, faster is the onset of aggregation of particles. $L=100, M_0=5,\epsilon=0.9, q_0=1, H=1500, v=2, C_1=0.2, C_2=0, D_R=1, \Gamma=0.5, \alpha_0=0.5, g=1, c_0=3.2, \sigma=25, \mathbf{r}_0=(50,50)$.}}
    \label{fig2}
\end{figure*}

We simulated Eq.~\ref{eq3.8} and Eq.~\ref{eq3.9} with an attractant concentration field given by the gaussian distribution $c(\mathbf{r},t)=c_0\exp^{-(\mathbf{r}-\mathbf{r}_0)/2\sigma^2}$, that remains constant in time. The strength of chemotactic interactions is specified by $c_0$, and the standard deviation by $\sigma$. We choose $\mathbf{r}_0$ to be the center of the simulation field. Fig.~\ref{fig2} shows snapshots of density and velocity profiles at four different time points and for three different densities, for a high value of chemotactic strength ($c_0=3.2$). The colormap in the background of each snapshot depicts the attractant concentration field. In each snapshot, particles are depicted by the solid red circles, centred around the density peaks of Eq.~\ref{eq3.8} at that point in time. Each particle's velocity is depicted by a yellow arrow starting from its center and pointing in the direction of the average polarization. The length of this arrow is a measure of the particle's speed. The particles with no arrows correspond to those that are halted and tumbling to find a new direction. For each of the three densities, we observe that initially the particles move randomly in all directions (Fig.~\ref{fig2} (a, e, i)), but with time, start climbing up the gradient of the attractant (Fig.~\ref{fig2} (b, f, j)). This results in the particles starting to accumulate in the center of the arena, which is the region of maximum attractant concentration (Fig.~\ref{fig2} (c, g, k)). Once there, these particles mostly tumble or move at low speeds, never moving very far away from the center. Eventually almost all particles aggregate around the center (Fig.~\ref{fig2} (d, h, l)). The higher the density the faster is this aggregation, due to the aligning effect of inelastic collisons discussed in section~\ref{sec2}. This aligning effect causes nearby particles to move in the same direction, allowing them to locate the upward gradient of the attractant faster than if each of them were searching independently (Fig.~\ref{fig2} (b, f, j)). 
\begin{figure*}
\begin{center}
\includegraphics[scale=0.2]{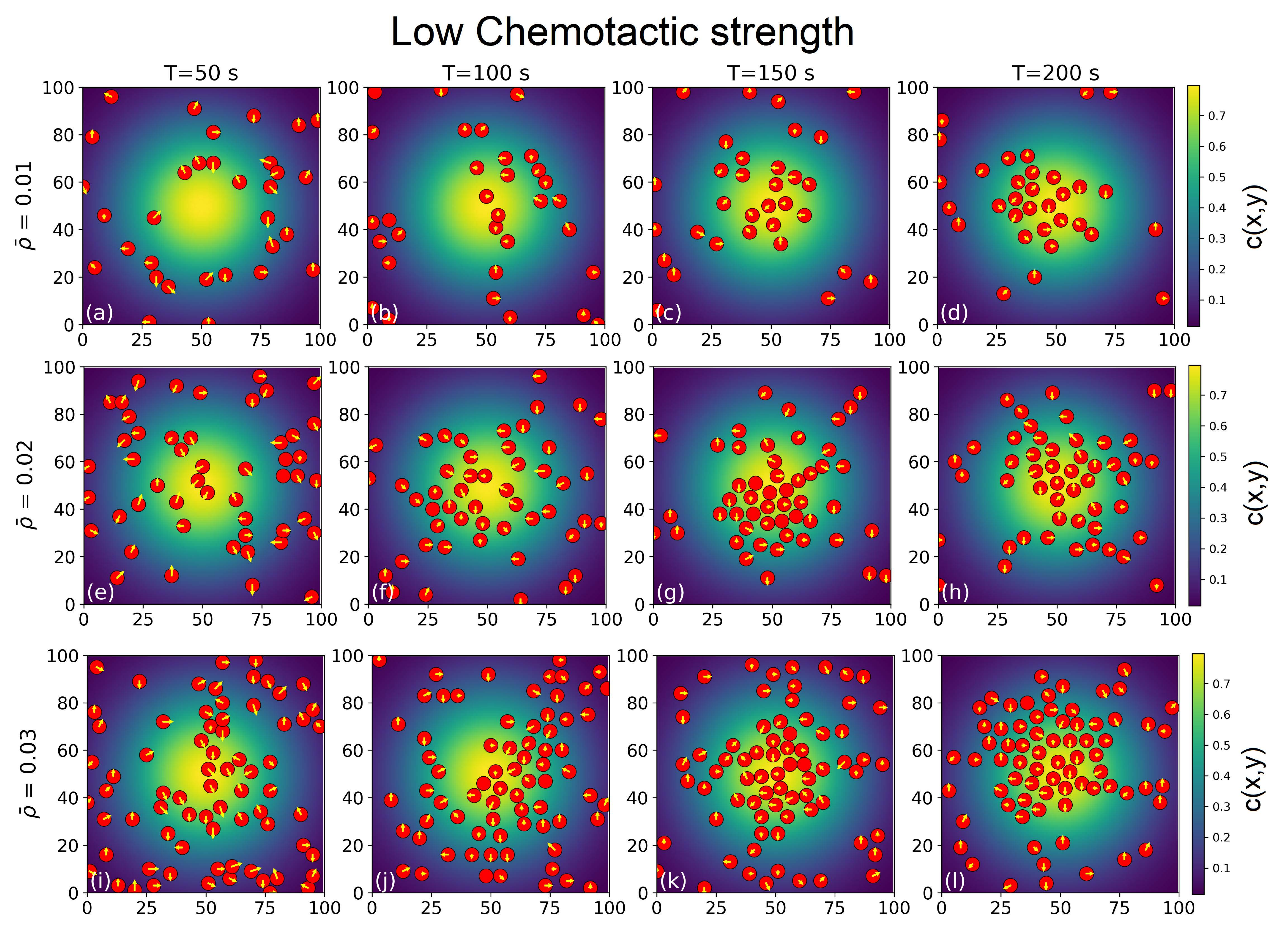}
\end{center}
\caption {\small{(Color online). Snapshots of density and velocity profiles at four different times $(50s, 100s, 150s, 200s)$, for average densities $\bar{\rho}=0.01$ (a, b, c, d), $\bar{\rho}=0.02$ (e, f, g, h), and $\bar{\rho}=0.03$ (i, j, k, l).  The chemical concentration field $c(x,y)$ is represented by the colormap. The chemotactic strength is low compared to that in Fig.~\ref{fig2}. This results in the observed lowering in the tendency of particles to aggregate.  $L=100, M_0=5,\epsilon=0.9, q_0=1, H=1500, v=2, C_1=0.2, C_2=0, D_R=1, \Gamma=0.5, \alpha_0=0.5, g=1, c_0=0.8, \sigma=25, \mathbf{r}_0=(50,50)$.}}
    \label{fig2b}
\end{figure*}

The strength of chemotactic interactions can be lowered by decreasing $c_0$, the maximum value of the attractant concentration. Fig.~\ref{fig2b} shows the result of lowering $c_0$ on the same three densities represented in Fig.~\ref{fig2}. It is clear from a comparison of Fig.~\ref{fig2} ($c_0=3.2$) and Fig.~\ref{fig2b} ($c_0=0.8$), that for lower chemotactic strengths, the tendency of particles to aggregate is much weaker. For $c_0=0.8$, there is very little aggregation at the center where the attractant field has the highest concentration, even after $200$ s of simulation time (Fig.~\ref{fig2b} (d, h, l)). This is qualitatively consistent with our schematic phase diagram (Fig.~\ref{fig0}), which predicts a decrease in aggregation with decreases chemotactic strength. We therefore conclude that the active VPFC chemotaxis model is able to capture the dynamics of run and tumble particles aggregating in a constant attractant field. Assuming that the time scale of this aggregation is much smaller than that of variations in the external stimulus, this is then an ideal framework to study thermotactic and phototactic particles.


\subsection{Self Chemotaxis}
\begin{figure*}
\begin{center}
\includegraphics[scale=0.2]{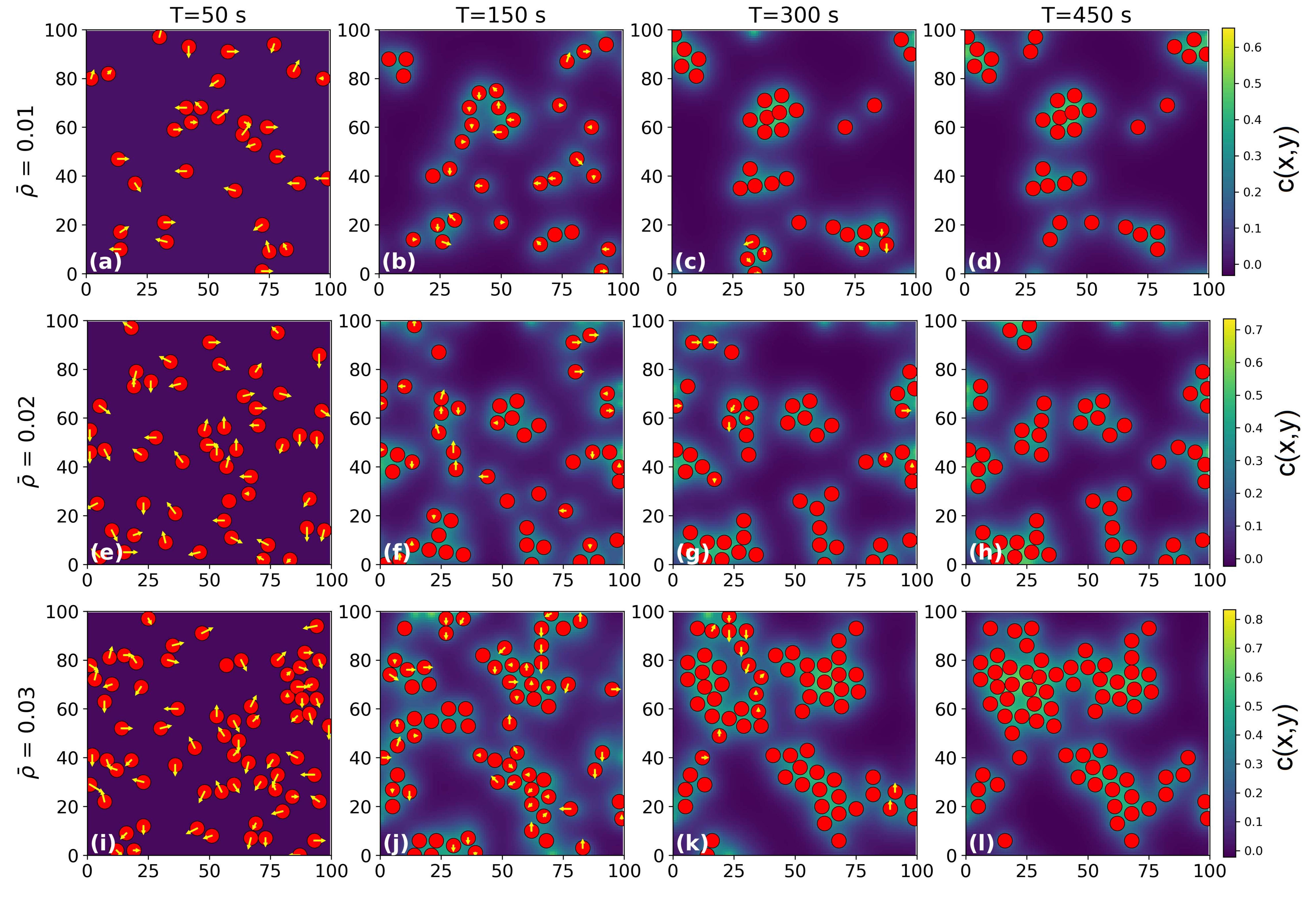}
\end{center}
\caption {\small{(Color online). Snapshots of the density profile at four different times $(50s, 150s, 300s, 450s)$, for average densities $\bar{\rho}=0.01$ (a, b, c, d), $\bar{\rho}=0.02$ (e, f, g, h), and $\bar{\rho}=0.03$ (i, j, k, l). The chemical concentration field $c(x,y)$ is represented by the colormap. Initially, particles move in random directions, but by $150s$ start forming clusters and comoving domains that allow them to collective search for high attractant regions. By $300$ most of the particles aggregate, with a only a few still searching. By $450s$, all particles come to rest in scattered aggregates. Higher the density, faster is the onset of aggregation of particles. $L=100, M_0=5,\epsilon=0.9, q_0=1, H=1500, v=2, C_1=0.2, C_2=0, D_R=D_c=1, \Gamma=0.5, \alpha_0=0.5, g=1, b=0.35, d_0=0.5.$}}
    \label{fig3}
\end{figure*}
We now consider the case of self-chemotaxis, where the attractant in question is secreted by the run and tumble particle itself. As a result the attractant concentration field varies with time, with higher values in regions of high particle density. We also allow the attractant to diffuse and degrade proportional to its own local concentration. The attractant field then evolves in time according to the formulation in the simplified Keller-Segel model,
\begin{equation}\label{eq3.2.1}
\partial_t c=D_c\nabla^2c+b\rho-d(c)c,
\end{equation}
where $D_c$ is the diffusion coefficient of the attractant, $b$ is the constant rate at it which it is produced by the particles, and $d(c)$ is the degradation rate of the attractant, that depends on its local concentration as $d(c)=d_0/(1+c)$. Higher production rates and lower degradation rates facilitate chemotaxis by generating larger gradients of the attractant field for the particle to climb up. Faster diffusion facilitates chemotaxis by increasing the range of one particles effect on another.  

\begin{figure*}
\begin{center}
\includegraphics[scale=0.25]{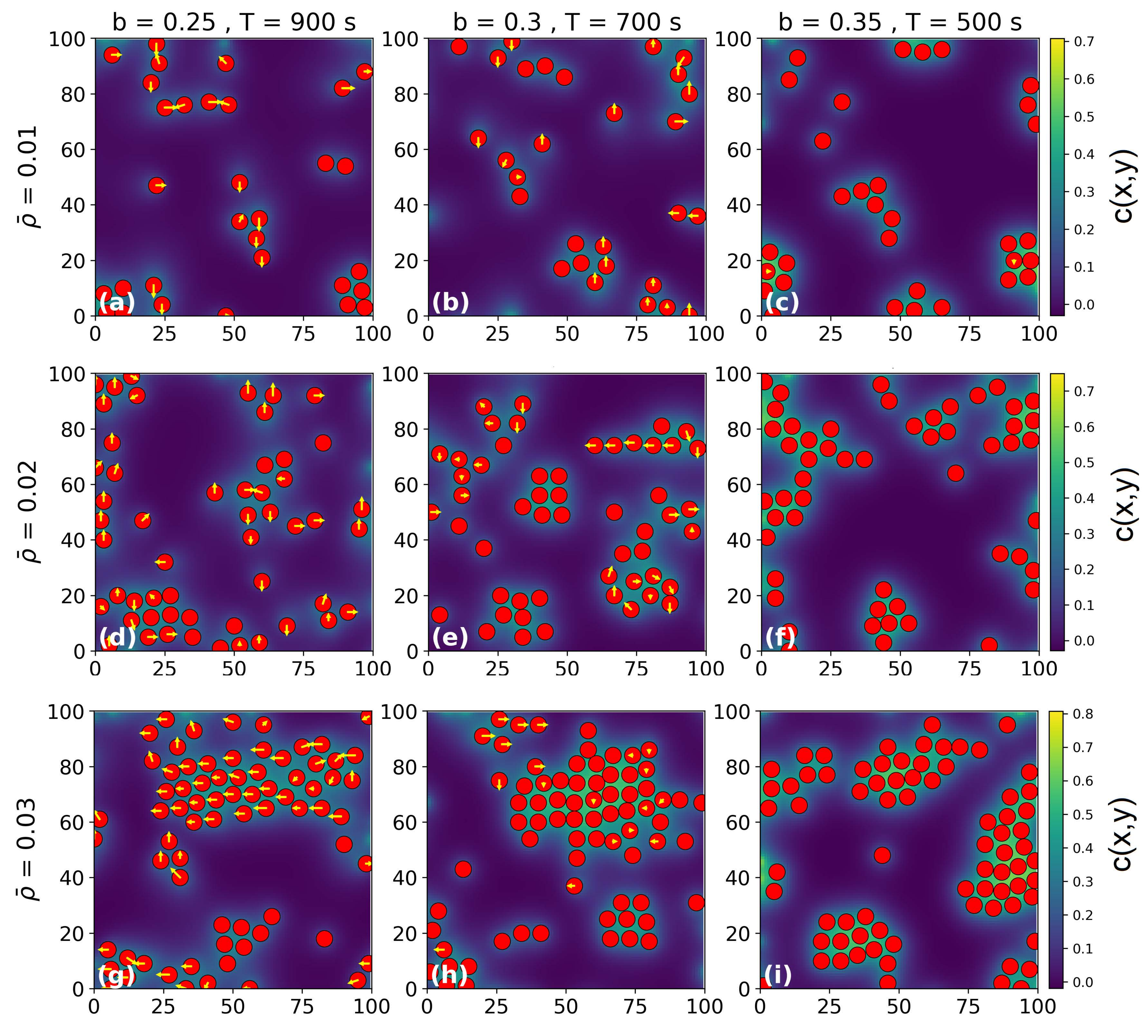}
\end{center}
\caption {\small{(Color online). Snapshots of the density profile for three different rates of attractant production $b=(0.25, 0.30, 0.35)$, for average densities $\bar{\rho}=0.03$ (panels (a),(b) and (c)), and $\bar{\rho}=0.06$ (panels (d),(e) and (f)). For $b=0.25$ particles do not settle into stationary aggregates even after $900s$ of simulation time. Instead, they form domains of comoving particles, with a few small clusters at rest. For $b=0.30$, most of the particles arrange themselves in stationary aggregates by $700s$, with only a few travelling domains. For $b=0.35$, all particles aggregate within $500$s of simulation time. $L=100, M_0=5,\epsilon=0.9, q_0=1, H=1500, v=2, C_1=0.2, C_2=0, D_R=D_c=1, \Gamma=0.5, \alpha_0=0.5, g=1, d_0=0.5$.}}
    \label{fig4}
\end{figure*}

Eq.~\ref{eq3.8} and Eq.~\ref{eq3.9}, along with Eq.~\ref{eq3.2.1} form the active VPFC self-chemotaxis model, which we simulate. Fig.~\ref{fig3} show snapshots of density profiles and particle velocities for three different average densities, and at four time points. Once again the particles are depicted by red solid circles and the velocity of each particle by a yellow arrow. Particles with no arrows are at rest and tumbling. The attractant field is provided by the colormap in the background. We observe that at $T=50s$, when the production of the attractant starts, the particles are moving randomly in all directions for all three densities (Fig.~\ref{fig3} (a, e, i)). By $T=150$s, some of the particles have begun to cluster and are halted,  while the others form domains of particles travelling in a common direction on average (Fig.~\ref{fig3} (b, f, j)). At $T=300s$, most of the travelling domains have collided, either with each other or with the stationary clusters, and come to a halt. The result is the appearance of aggregates of stationary particles, with only a small fraction of the particles still performing runs to find an upward gradient of the attractant (Fig.~\ref{fig3} (c, g, k)). Finally, by $450s$ all of the particles have come to a halt in scattered aggregates (Fig.~\ref{fig3} (d, h, l)). The number of such aggregates decrease with increasing density, with two to three stationary aggregates for the highest density represented (Fig.~\ref{fig3} (l)), and about six to eight aggregates for the lower densities represented (Fig.~\ref{fig3} (d, h)). This dependence of the equilibrium number of aggregates on density is once again a result of the cooperative aligning effect of inelastic collisions between particles, which forms domains of co-moving particles and allows them to collectively search for high attractant concentrations. In some of the snapshots at $T=450s$, we see some particles that are stationary but do not form a part of any aggregate (Fig.~\ref{fig3} (d,l)). These particles are isolated from nearby aggregates and are arrested in a state of perpetual tumbling due to the lack of the cooperative aligning effect of particles nearby. For the very same reason, the appearance of discernible stationary clusters occurs faster with increasing densities, as a comparison of the top, middle and bottom panels of Fig.~\ref{fig3} reveals. Regions of high particle density are also pockets of high average attractant concentration, with weaker gradients in its bulk as compared to its edges. As a result, as new particles approach an aggregate, their running speeds are progressively lowered till they come to a halt, reinforcing the aggregation process. Thus, even though we impose a constant mobility $M_0$ on the particles, a density dependent mobility emerges from the chemotactic interactions among individuals. Our conclusion is that even though particles aggregate to some extent at all densities, the probability of formation of large connected aggregates improves with increasing average densities. 

We next investigated the effect of the attractant secretion rate $b$ on the formation of these aggregates. Fig~\ref{fig4} shows snapshots of density and velocity profiles for the same three densities depicted in Fig.~\ref{fig3}, but for three different values of $b$. For $b=0.25$, particles do not settle into stationary aggregates even after $900s$ from the start of the simulation (Fig.~\ref{fig4} (a, d, g)). Instead we mostly observe domains of co-moving particles, and a very few small, stationary clusters. For $b=0.3$, stationary aggregates appear within $700s$, with a few remaining travelling domains (Fig.~\ref{fig4} (b, e, h)) that haven't settled into any aggregates. For $b=0.35$, which is the value that was used for Fig.~\ref{fig3}, all particles settle into 
aggregates by $500$s. Increasing rates of attractant secretion thus increases the strength of chemotactic activity, and assists in particle aggregation. This is again qualitatively consistent with our expected phase diagram (Fig.\ref{fig0}). 

We conclude that the active VPFC chemotaxis model captures the essential features of aggregation through self chemotaxis. Chemotactic aggregation has an important biological significance. A large number of cellular responses such as virulence and biofilm formation, require the coordinated action of many individual cells, and are triggered by the achievement of a threshold cell density, or ``quorum". Self chemotaxis has experimentally been proven to be an important mechanism employed by cells to generate the high cell densities required for quorum sensing \cite{Park2003,Park2003a}. The active VPFC self-chemotaxis model is thus the ideal framework to study these cellular responses on a diffusive timescale, but at the level of individual particle interactions.


\section{Discussion}\label{sec4}

In this paper we derived the active VPFC model from density functional theory, as a field-theoretical description of active particles. We showed that it is able to capture the long range orientational ordering of active particles, and the formation of traveling bands. The collective migration of particles is facilitated by inelastic collisions that deplete the transverse components of particle velocities, aligning them, and ensuring that no two particles can get closer to each other than their diameter. We then derived an active VPFC chemotaxis model for run and tumble particles, once again starting from density functional theory. We showed that this model is able to capture the migration of particles to regions of high attractant concentration through collective gradient sensing, once again facilitated by the aligning effect of inelastic collisions. We studied this model of chemotaxis for two different cases of attractant concentration dynamics. First, we showed that this model is able to capture the collective behavior of phototactic or thermotactic agents, which aggregate in regions of high levels of light or heat. Secondly, we showed that this model is also able to capture aggregation through self-chemotaxis in particles that themselves secrete the attractant. Such aggregation has been proposed as an important mechanism facilitating quorum sensing in bacteria. 

\begin{figure}
\begin{center}
\includegraphics[scale=0.25]{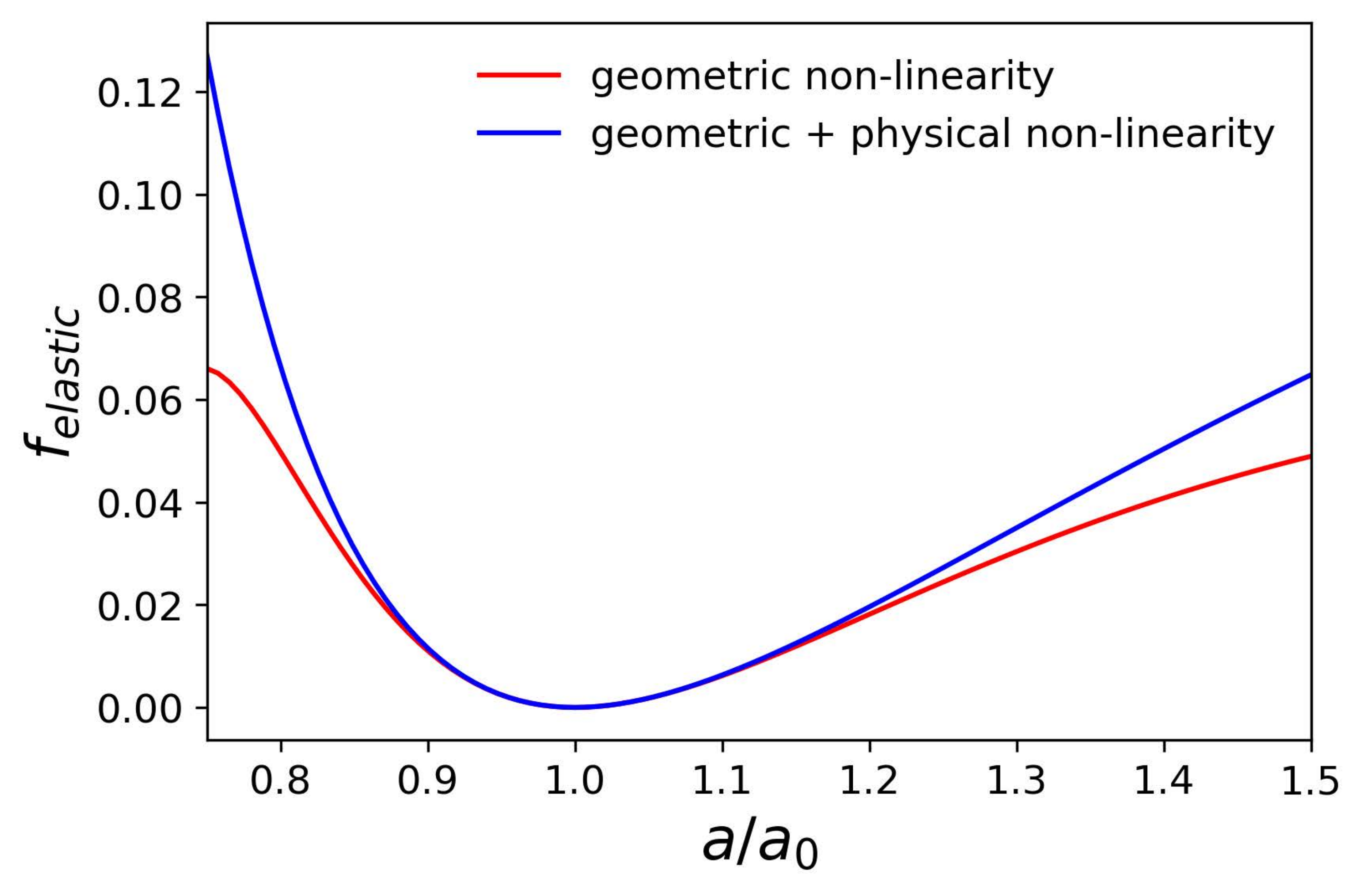}
\end{center}
\caption {\small{(Color online). Elastic energy density as a function of the lattice constant $a$ in the $1$D PFC model for $\bar{\rho}=0.3$ and $\epsilon=0.9$. The blue curve captures the geometric non-linearity for small deformations which arises because of nonlinear terms in the eulerian Alamansi strain tensor. The constitutive law relating the stress to the strain is however linear in this regime, since the elastic free energy is quadratic in the strain. The red curve and captures the physical non-linearity that becomes prominent for large deformations, when higher order terms in the strain cannot be neglected. For a derivation of the elastic free energy and its formulation in terms of the eulerian strain in both regimes, refer to the Appendix.}}
    \label{fig5}
\end{figure}

It is straightforward to generalize the active VPFC framework for the description of not just run and tumble particles, but of mobile aggregates and spatially patterned collective phases that emerge from an explicit density dependence of the particle mobility. Such phases have been observed for example in mussel beds, ant colonies and animal herds \cite{Liu2013,Theraulaz2002,Liu2016} and predicted by theoretical models of self-propelled rods \cite{Farrell2012}. However, as touched upon in the introduction, the most promising future application of the active VPFC framework lies in the investigation of the material properties of active systems. The equilibrium PFC model, which is the starting point for the active VPFC model, has already been shown to capture both linear and non-linear elasticity under bulk stress \cite{Chan2009,Huter2016}. This was done by looking for periodic solutions of the density field, and studying the increase in the free energy under isotropic strain due to deformation of the underlying lattice. Here it is important to note that the ground state in the equilibrium PFC case is perfectly crystalline (no vacancies and zero self-propulsion velocity) with an inverse lattice constant $q_0$, in contrast to the active VPFC ground state (vacancies and finite self-propulsion velocity). The authors of \cite{Chan2009,Huter2016} employ a one-mode expansion to describe the equilibrium PFC density field $\rho$ in $1$D,
\begin{equation}
\rho(x)=A(q) \cos(qx) +\bar{\rho},\label{eq4.1}
\end{equation}
where $q$ is the inverse lattice constant of the resulting periodic state and $A(q)$ is a strain dependent amplitude with $A(q_0)=A_0$. This ansatz considers the density field to be a superposition of plane waves, which for small values of the average density $\bar{\rho}$ and reduced temperature $\epsilon$ gives an excellent description of the true density \cite{Elder2002} in the equilibrium PFC model. Deviations of $q$ from $q_0$ increases the free energy, and allows us to calculate the elastic free energy density as $f_{elastic}(q)=f(q)-f(q_0)$, where $f(q)$ is the free energy density averaged over a unit cell $a=2\pi/q$. It was shown that both geometric as well as physical non-linearity in elastic response can be captured by the PFC model in not only one, but also in two and three dimensions. Fig.~\ref{fig5} plots the elastic free energy density as a function of the lattice constant $a$ of the deformed lattice in $1$D, for an average density $\bar{\rho}=0.3$ and reduced temperature $\epsilon=0.9$. We see that the non-linear elastic response is more stiff under compression ($a<a_0$), then under tension ($a>a_0$), as one would expect physically. A complete derivation of the elastic free energy (Fig.~\ref{fig5}) from the equilibrium PFC free energy is given in the Appendix.

In future work, we would like to apply the same technique to the study of elastic response in active systems. In any collective phase of SPPs that manifests as aggregation or cohesivity among particles, one can look look for solutions of the density field that are locally periodic. The challenge lies in the fact that the ground state is not perfectly crystalline in the case of active particles. The vacancy term in the active VPFC free energy, while it allows us to interpret $\rho$ as a physical, conserved density, also calls in to question whether a one-mode expansion such as Eq.~\ref{eq4.1} is still a good representation of the true density. However, we believe that the vacancy constraint can be imposed on the equilibrium PFC elastic theory through perturbative expansion, since we are most interested in the long wavelength limit. 

\setcounter{equation}{0}

\makeatletter
\renewcommand{\theequation}{A\arabic{equation}}
\renewcommand{\bibnumfmt}[1]{[A#1]}

\section*{Appendix}
Here we will go through the derivation of the elastic free-energy in the equilibrium PFC model, in order to explain the conceptual approach that we want to apply to active systems in future. This derivation ultimately gives us the elastic free energy density as a function of the lattice constant $a=2\pi/q$, which corresponds to Fig.~\ref{fig5}. We restrict ourselves to $1$D for simplicity, and a complete derivation of the elastic response in higher dimensions can be found in \cite{Huter2016,Chan2009}. 

The density field in the equilibrium PFC model is driven by the free energy functional $\mathcal{F}_{PFC}$, which penalizes deviations from periodicity and thus imposes a perfectly crystalline ground state \cite{Elder2002},
\begin{align}
\partial_t \rho&= \nabla^2\Big(\frac{\delta \mathcal{F}_{PFC}}{\delta \rho}\Big).\label{eq4.1}\\
\mathcal{F}_{PFC}[\rho]&=\int d\mathbf{r}\hspace{5pt}\Bigg[\frac{\rho}{2}\bigg[-\epsilon + (q_0^2+\nabla^2)^2\bigg]\rho +\frac{\rho^4}{4}\Bigg].\label{eq4.2}
\end{align}
The free energy $\mathcal{F}_{PFC}$ is the first term in the VPFC free energy functional $\mathcal{F}_{VPFC}$ given in Eq.~\ref{eq2.8}, with $\lambda$ set to one. It is therefore the equilibrium PFC free energy which imparts to the active VPFC model an underlying crystalline structure. A deformation field $\bm{\chi}$ displaces a material point with initial position $\mathbf{X}$ to its current position $\mathbf{x}$ as $\mathbf{x} =\bm{\chi}(\mathbf{X})$. As shown in \cite{Huter2016}, considering the energy density in an Eulerian sense allows for a useful physical interpretation of elasticity in the equilibrium PFC model. Thus we define the displacement field in the Eulerian sense as $\mathbf{u}(\mathbf{x})=\mathbf{x}-\bm{\chi}^{-1}(\mathbf{x})$. The deformation of the line element $d\mathbf{X}$ relative to $d\mathbf{x}$ in the current reference frame is given by 
\begin{equation}\label{eq4.3}
d\mathbf{X}=\nabla\bm{\chi}^{-1}(\mathbf{x}).d\mathbf{x}=\Big(\mathbf{I}-\nabla\mathbf{u}(\mathbf{x})\Big).d\mathbf{x}.
\end{equation}
The squared change of length of the line elements $ds^2=|d\mathbf{x}|^2-|d\mathbf{X}|^2$ can then be written as
$ds^2=dx_i e_{ij} dx_j$, where $e_{ij}$ is the Eulerian Alamansi strain tensor, defined in terms of the displacement field $\mathbf{u}$ in $N$ dimensions as,
\begin{equation}\label{eq4.4}
e_{ij}=\frac{1}{2}\Bigg(\frac{\partial u_i}{\partial x_j} + \frac{\partial u_j}{\partial x_i}-\frac{\partial u_m}{\partial x_i}\frac{\partial u_m}{\partial x_j}\Bigg).
\end{equation}
Here $u_i$ is the displacement in the $i^{th}$ direction, $i=1,2,...,N$, and Einstein summation is implied. In order to analytically calculate the elastic response of the crystalline ground state in $1$D, we employ a one-mode expansion of the density field,
\begin{equation}
\rho(x)=A(q) \cos(qx) +\bar{\rho},\label{eq4.5}
\end{equation}
where $q$ is the inverse lattice constant of the resulting periodic state and $A(q)$ is a strain dependent amplitude with $A(q_0)=A_0$. This ansatz considers the density field to be a superposition of plane waves, which for small values of the average density $\bar{\rho}$ and reduced temperature $\epsilon$ gives an excellent description of the true density \cite{Elder2002}. Inserting Eq.~\ref{eq4.5} into Eq.~\ref{eq4.2} and averaging over one unit cell $a=2\pi/q$, we obtain the free energy density of the system,
\begin{align}\label{eq4.6}
f&=\frac{1}{a}\int_0^a dx \hspace{5pt} \frac{\delta \mathcal{F}_{PFC}}{\delta x},\nonumber\\
&= \frac{\bar{\rho}^2}{2}\Big(-\epsilon+q_0^2+\frac{3A^2(q)}{2}+\frac{\bar{\rho}^2}{2}\Big)\nonumber\\
&+\frac{A^2(q)}{4}\Big(-\epsilon+(q_0^2-q^2)^2+\frac{3A^2(q)}{8}\Big).
\end{align}
The strain dependent amplitude $A(q)$ is chosen such that it minimizes the free energy density $f$ in Eq.~\ref{eq4.6} for a given $\epsilon$ and $\bar{\rho}$. This gives us
\begin{equation}\label{eq4.7}
A^2(q)=\frac{4}{3}\Big(\epsilon-3\bar{\rho}^2-(q_0^2-q^2)^2\Big).
\end{equation}
Inserting Eq.~\ref{eq4.7} into the free energy density Eq.~\ref{eq4.6}, we get the final free energy density
\begin{align}\label{eq4.8}
f&=\frac{7\bar{\rho}^4}{4}+\frac{\bar{\rho}^2}{2}\Big(\epsilon+q_0^2-2(q_0^2-q^2)^2\Big)\nonumber\\
&-\frac{1}{6}\Big((q_0^2-q^2)^2-\epsilon\Big)^2.
\end{align}
The free energy density is minimized for $q=q_0$ for a given $\epsilon$ and $\bar{\rho}$. Deviations in $q$ from $q_0$imposes a homogeneous strain in the system and causes the free energy to increase. In $1$D, the displacement vector is given by 
\begin{equation}\label{eq4.9}
u(x)=(q_0-q)x.
\end{equation}
The Eulerian strain tensor in $1$D, according to Eq.~\ref{eq4.4} is then 
\begin{equation}\label{eq4.10}
e_{xx}=\frac{1}{2}\Big(q_0^2-q^2\Big).
\end{equation}
We can thus write down the elastic free energy density $f_{elastic}(q)=f(q)-f(q_0)$ in terms of the strain tensor $e_{xx}$ as,
\begin{equation}\label{eq4.11}
f_{elastic}=e_{xx}^2\Big(\frac{4\epsilon}{3}-4\bar{\rho}^2\Big)-\frac{8e_{xx}^4}{3}.
\end{equation}
For small deviations of $q$ from $q_0$, we can ignore the fourth order term in $e_{xx}$ in the elastic free energy, and get
\begin{equation}\label{eq4.12}
f_{elastic}=A_0^2e_{xx}^2.
\end{equation}
This tells us that even though the elastic free energy is not a quadratic function of the displacement gradient $u(x)$, this non-linearity is purely geometric in nature and arises because of nonlinear terms in the strain tensor. The constitutive law relating the stress to the strain is however linear in this regime, since the elastic free energy is quadratic in the strain $e_{xx}$. For larger strains the fourth order term in $e_{xx}$ becomes important, and the non-linearity is both geometric and physical in nature. Fig.~\ref{fig5} illustrates this for $\bar{\rho}=0.3$ and $\epsilon=0.9$, by sketching the elastic free energy for both regimes as a function of $a=2\pi/q$. We see that the non-linear elastic response is more stiff under compression ($a<a_0$), then under tension ($a>a_0$), as one would expect physically.

\bibliographystyle{apsrev4-2}
\bibliography{active_VPFC_chemotaxis}

\begin{thebibliography}{45}%
\makeatletter
\providecommand \@ifxundefined [1]{%
 \@ifx{#1\undefined}
}%
\providecommand \@ifnum [1]{%
 \ifnum #1\expandafter \@firstoftwo
 \else \expandafter \@secondoftwo
 \fi
}%
\providecommand \@ifx [1]{%
 \ifx #1\expandafter \@firstoftwo
 \else \expandafter \@secondoftwo
 \fi
}%
\providecommand \natexlab [1]{#1}%
\providecommand \enquote  [1]{``#1''}%
\providecommand \bibnamefont  [1]{#1}%
\providecommand \bibfnamefont [1]{#1}%
\providecommand \citenamefont [1]{#1}%
\providecommand \href@noop [0]{\@secondoftwo}%
\providecommand \href [0]{\begingroup \@sanitize@url \@href}%
\providecommand \@href[1]{\@@startlink{#1}\@@href}%
\providecommand \@@href[1]{\endgroup#1\@@endlink}%
\providecommand \@sanitize@url [0]{\catcode `\\12\catcode `\$12\catcode
  `\&12\catcode `\#12\catcode `\^12\catcode `\_12\catcode `\%12\relax}%
\providecommand \@@startlink[1]{}%
\providecommand \@@endlink[0]{}%
\providecommand \url  [0]{\begingroup\@sanitize@url \@url }%
\providecommand \@url [1]{\endgroup\@href {#1}{\urlprefix }}%
\providecommand \urlprefix  [0]{URL }%
\providecommand \Eprint [0]{\href }%
\providecommand \doibase [0]{https://doi.org/}%
\providecommand \selectlanguage [0]{\@gobble}%
\providecommand \bibinfo  [0]{\@secondoftwo}%
\providecommand \bibfield  [0]{\@secondoftwo}%
\providecommand \translation [1]{[#1]}%
\providecommand \BibitemOpen [0]{}%
\providecommand \bibitemStop [0]{}%
\providecommand \bibitemNoStop [0]{.\EOS\space}%
\providecommand \EOS [0]{\spacefactor3000\relax}%
\providecommand \BibitemShut  [1]{\csname bibitem#1\endcsname}%
\let\auto@bib@innerbib\@empty
\bibitem [{\citenamefont {Keller}\ and\ \citenamefont
  {Segel}(1970)}]{Keller1970}%
  \BibitemOpen
  \bibfield  {author} {\bibinfo {author} {\bibfnamefont {E.~F.}\ \bibnamefont
  {Keller}}\ and\ \bibinfo {author} {\bibfnamefont {L.~A.}\ \bibnamefont
  {Segel}},\ }\href@noop {} {\bibfield  {journal} {\bibinfo  {journal} {Journal
  of Theoretical Biology}\ } (\bibinfo {year} {1970})}\BibitemShut {NoStop}%
\bibitem [{\citenamefont {Adler}(1966)}]{Adler1966}%
  \BibitemOpen
  \bibfield  {author} {\bibinfo {author} {\bibfnamefont {J.}~\bibnamefont
  {Adler}},\ }\href@noop {} {\bibfield  {journal} {\bibinfo  {journal}
  {Science}\ } (\bibinfo {year} {1966})}\BibitemShut {NoStop}%
\bibitem [{\citenamefont {Budrene}\ and\ \citenamefont
  {Berg}(1995)}]{Budrene1995}%
  \BibitemOpen
  \bibfield  {author} {\bibinfo {author} {\bibfnamefont {E.~O.}\ \bibnamefont
  {Budrene}}\ and\ \bibinfo {author} {\bibfnamefont {H.~C.}\ \bibnamefont
  {Berg}},\ }\href@noop {} {\bibfield  {journal} {\bibinfo  {journal} {Nature}\
  } (\bibinfo {year} {1995})}\BibitemShut {NoStop}%
\bibitem [{\citenamefont {Maeda}\ \emph {et~al.}(1976)\citenamefont {Maeda},
  \citenamefont {Imae},\ and\ \citenamefont {J.~I. Shioi~and}}]{Maeda1976}%
  \BibitemOpen
  \bibfield  {author} {\bibinfo {author} {\bibfnamefont {K.}~\bibnamefont
  {Maeda}}, \bibinfo {author} {\bibfnamefont {Y.}~\bibnamefont {Imae}},\ and\
  \bibinfo {author} {\bibfnamefont {F.~O.}\ \bibnamefont {J.~I. Shioi~and}},\
  }\href@noop {} {\bibfield  {journal} {\bibinfo  {journal} {Journal of
  Bacteriology}\ } (\bibinfo {year} {1976})}\BibitemShut {NoStop}%
\bibitem [{\citenamefont {Mittal}\ \emph {et~al.}(2003)\citenamefont {Mittal},
  \citenamefont {Budrene}, \citenamefont {Brenner},\ and\ \citenamefont {van
  Oudenaarden}}]{Mittal2003}%
  \BibitemOpen
  \bibfield  {author} {\bibinfo {author} {\bibfnamefont {N.}~\bibnamefont
  {Mittal}}, \bibinfo {author} {\bibfnamefont {E.~O.}\ \bibnamefont {Budrene}},
  \bibinfo {author} {\bibfnamefont {M.~P.}\ \bibnamefont {Brenner}},\ and\
  \bibinfo {author} {\bibfnamefont {A.}~\bibnamefont {van Oudenaarden}},\
  }\href@noop {} {\bibfield  {journal} {\bibinfo  {journal} {PNAS}\ } (\bibinfo
  {year} {2003})}\BibitemShut {NoStop}%
\bibitem [{\citenamefont {Park}\ \emph
  {et~al.}(2003{\natexlab{a}})\citenamefont {Park}, \citenamefont {Wolanin},
  \citenamefont {Yuzbashyan}, \citenamefont {Silberzan}, \citenamefont
  {Stock},\ and\ \citenamefont {Austin}}]{Park2003}%
  \BibitemOpen
  \bibfield  {author} {\bibinfo {author} {\bibfnamefont {S.}~\bibnamefont
  {Park}}, \bibinfo {author} {\bibfnamefont {P.~M.}\ \bibnamefont {Wolanin}},
  \bibinfo {author} {\bibfnamefont {E.~A.}\ \bibnamefont {Yuzbashyan}},
  \bibinfo {author} {\bibfnamefont {P.}~\bibnamefont {Silberzan}}, \bibinfo
  {author} {\bibfnamefont {J.~B.}\ \bibnamefont {Stock}},\ and\ \bibinfo
  {author} {\bibfnamefont {R.~H.}\ \bibnamefont {Austin}},\ }\href@noop {}
  {\bibfield  {journal} {\bibinfo  {journal} {Science}\ } (\bibinfo {year}
  {2003}{\natexlab{a}})}\BibitemShut {NoStop}%
\bibitem [{\citenamefont {Park}\ \emph
  {et~al.}(2003{\natexlab{b}})\citenamefont {Park}, \citenamefont {Wolanin},
  \citenamefont {Yuzbashyan}, \citenamefont {Lin}, \citenamefont {Darnton},
  \citenamefont {Stock}, \citenamefont {Silberzan},\ and\ \citenamefont
  {Austin}}]{Park2003a}%
  \BibitemOpen
  \bibfield  {author} {\bibinfo {author} {\bibfnamefont {S.}~\bibnamefont
  {Park}}, \bibinfo {author} {\bibfnamefont {P.~M.}\ \bibnamefont {Wolanin}},
  \bibinfo {author} {\bibfnamefont {E.~A.}\ \bibnamefont {Yuzbashyan}},
  \bibinfo {author} {\bibfnamefont {H.}~\bibnamefont {Lin}}, \bibinfo {author}
  {\bibfnamefont {N.~C.}\ \bibnamefont {Darnton}}, \bibinfo {author}
  {\bibfnamefont {J.~B.}\ \bibnamefont {Stock}}, \bibinfo {author}
  {\bibfnamefont {P.}~\bibnamefont {Silberzan}},\ and\ \bibinfo {author}
  {\bibfnamefont {R.}~\bibnamefont {Austin}},\ }\href@noop {} {\bibfield
  {journal} {\bibinfo  {journal} {PNAS}\ } (\bibinfo {year}
  {2003}{\natexlab{b}})}\BibitemShut {NoStop}%
\bibitem [{\citenamefont {Budrene}\ and\ \citenamefont
  {Berg}(1991)}]{Budrene1991}%
  \BibitemOpen
  \bibfield  {author} {\bibinfo {author} {\bibfnamefont {E.~O.}\ \bibnamefont
  {Budrene}}\ and\ \bibinfo {author} {\bibfnamefont {H.~C.}\ \bibnamefont
  {Berg}},\ }\href@noop {} {\bibfield  {journal} {\bibinfo  {journal} {Nature}\
  } (\bibinfo {year} {1991})}\BibitemShut {NoStop}%
\bibitem [{\citenamefont {Woodward}\ \emph {et~al.}(1995)\citenamefont
  {Woodward}, \citenamefont {Tyson}, \citenamefont {Myerscough}, \citenamefont
  {Murray}, \citenamefont {Budrene},\ and\ \citenamefont
  {Berg}}]{Woodward1995}%
  \BibitemOpen
  \bibfield  {author} {\bibinfo {author} {\bibfnamefont {D.~E.}\ \bibnamefont
  {Woodward}}, \bibinfo {author} {\bibfnamefont {R.}~\bibnamefont {Tyson}},
  \bibinfo {author} {\bibfnamefont {M.~R.}\ \bibnamefont {Myerscough}},
  \bibinfo {author} {\bibfnamefont {J.~D.}\ \bibnamefont {Murray}}, \bibinfo
  {author} {\bibfnamefont {E.~O.}\ \bibnamefont {Budrene}},\ and\ \bibinfo
  {author} {\bibfnamefont {H.~C.}\ \bibnamefont {Berg}},\ }\href@noop {}
  {\bibfield  {journal} {\bibinfo  {journal} {Biophysical Journal}\ } (\bibinfo
  {year} {1995})}\BibitemShut {NoStop}%
\bibitem [{\citenamefont {Saragosti}\ \emph {et~al.}(2011)\citenamefont
  {Saragosti}, \citenamefont {Calvez}, \citenamefont {Bournaveas},
  \citenamefont {Perthame}, \citenamefont {Buguin},\ and\ \citenamefont
  {Silberzan}}]{Saragosti2011}%
  \BibitemOpen
  \bibfield  {author} {\bibinfo {author} {\bibfnamefont {J.}~\bibnamefont
  {Saragosti}}, \bibinfo {author} {\bibfnamefont {V.}~\bibnamefont {Calvez}},
  \bibinfo {author} {\bibfnamefont {N.}~\bibnamefont {Bournaveas}}, \bibinfo
  {author} {\bibfnamefont {B.}~\bibnamefont {Perthame}}, \bibinfo {author}
  {\bibfnamefont {A.}~\bibnamefont {Buguin}},\ and\ \bibinfo {author}
  {\bibfnamefont {P.}~\bibnamefont {Silberzan}},\ }\href@noop {} {\bibfield
  {journal} {\bibinfo  {journal} {PNAS}\ } (\bibinfo {year}
  {2011})}\BibitemShut {NoStop}%
\bibitem [{\citenamefont {Alaimo}\ \emph {et~al.}(2016)\citenamefont {Alaimo},
  \citenamefont {Praetorius},\ and\ \citenamefont {Voigt}}]{Alaimo2016}%
  \BibitemOpen
  \bibfield  {author} {\bibinfo {author} {\bibfnamefont {F.}~\bibnamefont
  {Alaimo}}, \bibinfo {author} {\bibfnamefont {S.}~\bibnamefont {Praetorius}},\
  and\ \bibinfo {author} {\bibfnamefont {A.}~\bibnamefont {Voigt}},\
  }\href@noop {} {\bibfield  {journal} {\bibinfo  {journal} {New J. Phys. 18
  083008}\ } (\bibinfo {year} {2016})}\BibitemShut {NoStop}%
\bibitem [{\citenamefont {Menzel}\ \emph {et~al.}(2014)\citenamefont {Menzel},
  \citenamefont {Ohta},\ and\ \citenamefont {L{\"o}wen}}]{Menzel2014}%
  \BibitemOpen
  \bibfield  {author} {\bibinfo {author} {\bibfnamefont {A.~M.}\ \bibnamefont
  {Menzel}}, \bibinfo {author} {\bibfnamefont {T.}~\bibnamefont {Ohta}},\ and\
  \bibinfo {author} {\bibfnamefont {H.}~\bibnamefont {L{\"o}wen}},\ }\href@noop
  {} {\bibfield  {journal} {\bibinfo  {journal} {Phys. Rev. E 89, 022301}\ }
  (\bibinfo {year} {2014})}\BibitemShut {NoStop}%
\bibitem [{\citenamefont {Suzuki}\ \emph {et~al.}(2015)\citenamefont {Suzuki},
  \citenamefont {Webera}, \citenamefont {Frey},\ and\ \citenamefont
  {Bausch}}]{Suzuki2015}%
  \BibitemOpen
  \bibfield  {author} {\bibinfo {author} {\bibfnamefont {R.}~\bibnamefont
  {Suzuki}}, \bibinfo {author} {\bibfnamefont {C.}~\bibnamefont {Webera}},
  \bibinfo {author} {\bibfnamefont {E.}~\bibnamefont {Frey}},\ and\ \bibinfo
  {author} {\bibfnamefont {A.}~\bibnamefont {Bausch}},\ }\href@noop {}
  {\bibfield  {journal} {\bibinfo  {journal} {Nature Physics volume 11, pages
  839–843}\ } (\bibinfo {year} {2015})}\BibitemShut {NoStop}%
\bibitem [{\citenamefont {Chatterjee}\ and\ \citenamefont
  {Goldenfeld}(2019)}]{Chatterjee2019}%
  \BibitemOpen
  \bibfield  {author} {\bibinfo {author} {\bibfnamefont {P.}~\bibnamefont
  {Chatterjee}}\ and\ \bibinfo {author} {\bibfnamefont {N.}~\bibnamefont
  {Goldenfeld}},\ }\href@noop {} {\bibfield  {journal} {\bibinfo  {journal}
  {Phys. Rev. E 100, 040602(R)}\ } (\bibinfo {year} {2019})}\BibitemShut
  {NoStop}%
\bibitem [{\citenamefont {Kelley}\ and\ \citenamefont
  {Ouellette}(2013)}]{Kelley2013}%
  \BibitemOpen
  \bibfield  {author} {\bibinfo {author} {\bibfnamefont {D.~H.}\ \bibnamefont
  {Kelley}}\ and\ \bibinfo {author} {\bibfnamefont {N.~T.}\ \bibnamefont
  {Ouellette}},\ }\href@noop {} {\bibfield  {journal} {\bibinfo  {journal}
  {Scientific Reports}\ } (\bibinfo {year} {2013})}\BibitemShut {NoStop}%
\bibitem [{\citenamefont {Puckett}\ and\ \citenamefont
  {Ouellette}(2014)}]{Puckett2014}%
  \BibitemOpen
  \bibfield  {author} {\bibinfo {author} {\bibfnamefont {J.~G.}\ \bibnamefont
  {Puckett}}\ and\ \bibinfo {author} {\bibfnamefont {N.~T.}\ \bibnamefont
  {Ouellette}},\ }\href@noop {} {\bibfield  {journal} {\bibinfo  {journal} {J.
  R. Soc. Interface 11: 20140710}\ } (\bibinfo {year} {2014})}\BibitemShut
  {NoStop}%
\bibitem [{\citenamefont {Puckett}\ \emph {et~al.}(2014)\citenamefont
  {Puckett}, \citenamefont {Kelley},\ and\ \citenamefont
  {Ouellette}}]{Puckett2014a}%
  \BibitemOpen
  \bibfield  {author} {\bibinfo {author} {\bibfnamefont {J.~G.}\ \bibnamefont
  {Puckett}}, \bibinfo {author} {\bibfnamefont {D.~H.}\ \bibnamefont
  {Kelley}},\ and\ \bibinfo {author} {\bibfnamefont {N.~T.}\ \bibnamefont
  {Ouellette}},\ }\href@noop {} {\bibfield  {journal} {\bibinfo  {journal}
  {Sci. Rep. 4, 4766}\ } (\bibinfo {year} {2014})}\BibitemShut {NoStop}%
\bibitem [{\citenamefont {Okubo}(1986)}]{Okubo1986}%
  \BibitemOpen
  \bibfield  {author} {\bibinfo {author} {\bibfnamefont {A.}~\bibnamefont
  {Okubo}},\ }\href@noop {} {\bibfield  {journal} {\bibinfo  {journal}
  {Advances in Biophysics}\ } (\bibinfo {year} {1986})}\BibitemShut {NoStop}%
\bibitem [{\citenamefont {Gorbonos}\ \emph {et~al.}(2016)\citenamefont
  {Gorbonos}, \citenamefont {Ianconescu}, \citenamefont {Puckett},
  \citenamefont {Ni}, \citenamefont {Ouellette},\ and\ \citenamefont
  {Gov}}]{Gorbonos2016}%
  \BibitemOpen
  \bibfield  {author} {\bibinfo {author} {\bibfnamefont {D.}~\bibnamefont
  {Gorbonos}}, \bibinfo {author} {\bibfnamefont {R.}~\bibnamefont
  {Ianconescu}}, \bibinfo {author} {\bibfnamefont {J.~G.}\ \bibnamefont
  {Puckett}}, \bibinfo {author} {\bibfnamefont {R.}~\bibnamefont {Ni}},
  \bibinfo {author} {\bibfnamefont {N.~T.}\ \bibnamefont {Ouellette}},\ and\
  \bibinfo {author} {\bibfnamefont {N.~S.}\ \bibnamefont {Gov}},\ }\href@noop
  {} {\bibfield  {journal} {\bibinfo  {journal} {New J. Phys. 18, 073042}\ }
  (\bibinfo {year} {2016})}\BibitemShut {NoStop}%
\bibitem [{\citenamefont {Chavanis}\ and\ \citenamefont
  {Sire}(2007)}]{Chavanis2007}%
  \BibitemOpen
  \bibfield  {author} {\bibinfo {author} {\bibfnamefont {P.~H.}\ \bibnamefont
  {Chavanis}}\ and\ \bibinfo {author} {\bibfnamefont {C.}~\bibnamefont
  {Sire}},\ }\href@noop {} {\bibfield  {journal} {\bibinfo  {journal} {Physica
  A 384 199–222}\ } (\bibinfo {year} {2007})}\BibitemShut {NoStop}%
\bibitem [{\citenamefont {Puckett}\ \emph {et~al.}(2015)\citenamefont
  {Puckett}, \citenamefont {Ni},\ and\ \citenamefont
  {Ouellette}}]{Puckett2015}%
  \BibitemOpen
  \bibfield  {author} {\bibinfo {author} {\bibfnamefont {J.~G.}\ \bibnamefont
  {Puckett}}, \bibinfo {author} {\bibfnamefont {R.}~\bibnamefont {Ni}},\ and\
  \bibinfo {author} {\bibfnamefont {N.~T.}\ \bibnamefont {Ouellette}},\
  }\href@noop {} {\bibfield  {journal} {\bibinfo  {journal} {Phys. Rev. Lett.
  114, 258103}\ } (\bibinfo {year} {2015})}\BibitemShut {NoStop}%
\bibitem [{\citenamefont {Reynolds}(2019{\natexlab{a}})}]{Reynolds2019}%
  \BibitemOpen
  \bibfield  {author} {\bibinfo {author} {\bibfnamefont {A.~M.}\ \bibnamefont
  {Reynolds}},\ }\href@noop {} {\bibfield  {journal} {\bibinfo  {journal} {J.
  R. Soc. Interface 16: 20190404}\ } (\bibinfo {year}
  {2019}{\natexlab{a}})}\BibitemShut {NoStop}%
\bibitem [{\citenamefont {Ni}\ and\ \citenamefont {Ouellette}(2016)}]{Ni2016}%
  \BibitemOpen
  \bibfield  {author} {\bibinfo {author} {\bibfnamefont {R.}~\bibnamefont
  {Ni}}\ and\ \bibinfo {author} {\bibfnamefont {N.~T.}\ \bibnamefont
  {Ouellette}},\ }\href@noop {} {\bibfield  {journal} {\bibinfo  {journal}
  {Phys. Biol. 13 045002}\ } (\bibinfo {year} {2016})}\BibitemShut {NoStop}%
\bibitem [{\citenamefont {van~der Vaart}\ \emph {et~al.}(2019)\citenamefont
  {van~der Vaart}, \citenamefont {Sinhuber}, \citenamefont {Reynolds},\ and\
  \citenamefont {T.Ouellette}}]{Vaart2019}%
  \BibitemOpen
  \bibfield  {author} {\bibinfo {author} {\bibfnamefont {K.}~\bibnamefont
  {van~der Vaart}}, \bibinfo {author} {\bibfnamefont {M.}~\bibnamefont
  {Sinhuber}}, \bibinfo {author} {\bibfnamefont {A.~M.}\ \bibnamefont
  {Reynolds}},\ and\ \bibinfo {author} {\bibfnamefont {N.}~\bibnamefont
  {T.Ouellette}},\ }\href@noop {} {\bibfield  {journal} {\bibinfo  {journal}
  {Sci. Adv. 5, eaaw9305}\ } (\bibinfo {year} {2019})}\BibitemShut {NoStop}%
\bibitem [{\citenamefont {Reynolds}(2019{\natexlab{b}})}]{Reynolds2019a}%
  \BibitemOpen
  \bibfield  {author} {\bibinfo {author} {\bibfnamefont {A.~M.}\ \bibnamefont
  {Reynolds}},\ }\href@noop {} {\bibfield  {journal} {\bibinfo  {journal}
  {Phys. Biol. 16 046002}\ } (\bibinfo {year}
  {2019}{\natexlab{b}})}\BibitemShut {NoStop}%
\bibitem [{\citenamefont {Pokhrel}\ \emph {et~al.}(2019)\citenamefont
  {Pokhrel}, \citenamefont {Kayastha},\ and\ \citenamefont
  {Puckett}}]{Pokhrel2019}%
  \BibitemOpen
  \bibfield  {author} {\bibinfo {author} {\bibfnamefont {A.}~\bibnamefont
  {Pokhrel}}, \bibinfo {author} {\bibfnamefont {P.}~\bibnamefont {Kayastha}},\
  and\ \bibinfo {author} {\bibfnamefont {J.}~\bibnamefont {Puckett}},\
  }\href@noop {} {\bibfield  {journal} {\bibinfo  {journal} {Bulletin of the
  American Physical Society}\ } (\bibinfo {year} {2019})}\BibitemShut {NoStop}%
\bibitem [{\citenamefont {Elder}\ \emph {et~al.}(2002)\citenamefont {Elder},
  \citenamefont {Katakowski}, \citenamefont {Haataja},\ and\ \citenamefont
  {Grant}}]{Elder2002}%
  \BibitemOpen
  \bibfield  {author} {\bibinfo {author} {\bibfnamefont {K.~R.}\ \bibnamefont
  {Elder}}, \bibinfo {author} {\bibfnamefont {M.}~\bibnamefont {Katakowski}},
  \bibinfo {author} {\bibfnamefont {M.}~\bibnamefont {Haataja}},\ and\ \bibinfo
  {author} {\bibfnamefont {M.}~\bibnamefont {Grant}},\ }\href@noop {}
  {\bibfield  {journal} {\bibinfo  {journal} {Phys. Rev. Lett. 88, 245701}\ }
  (\bibinfo {year} {2002})}\BibitemShut {NoStop}%
\bibitem [{\citenamefont {Chan}\ and\ \citenamefont
  {Goldenfeld}(2009)}]{Chan2009}%
  \BibitemOpen
  \bibfield  {author} {\bibinfo {author} {\bibfnamefont {P.}~\bibnamefont
  {Chan}}\ and\ \bibinfo {author} {\bibfnamefont {N.}~\bibnamefont
  {Goldenfeld}},\ }\href@noop {} {\bibfield  {journal} {\bibinfo  {journal}
  {Phys. Rev. E 80, 065105(R)}\ } (\bibinfo {year} {2009})}\BibitemShut
  {NoStop}%
\bibitem [{\citenamefont {Huter}\ \emph {et~al.}(2016)\citenamefont {Huter},
  \citenamefont {Friak}, \citenamefont {Weikamp}, \citenamefont {Neugebauer},
  \citenamefont {Goldenfeld}, \citenamefont {Svendsen},\ and\ \citenamefont
  {Spatschek}}]{Huter2016}%
  \BibitemOpen
  \bibfield  {author} {\bibinfo {author} {\bibfnamefont {C.}~\bibnamefont
  {Huter}}, \bibinfo {author} {\bibfnamefont {M.}~\bibnamefont {Friak}},
  \bibinfo {author} {\bibfnamefont {M.}~\bibnamefont {Weikamp}}, \bibinfo
  {author} {\bibfnamefont {J.}~\bibnamefont {Neugebauer}}, \bibinfo {author}
  {\bibfnamefont {N.}~\bibnamefont {Goldenfeld}}, \bibinfo {author}
  {\bibfnamefont {B.}~\bibnamefont {Svendsen}},\ and\ \bibinfo {author}
  {\bibfnamefont {R.}~\bibnamefont {Spatschek}},\ }\href@noop {} {\bibfield
  {journal} {\bibinfo  {journal} {Phys. Rev. B 93, 214105}\ } (\bibinfo {year}
  {2016})}\BibitemShut {NoStop}%
\bibitem [{\citenamefont {Chan}\ \emph {et~al.}(2009)\citenamefont {Chan},
  \citenamefont {Goldenfeld},\ and\ \citenamefont {Dantzig}}]{Chan2009a}%
  \BibitemOpen
  \bibfield  {author} {\bibinfo {author} {\bibfnamefont {P.~Y.}\ \bibnamefont
  {Chan}}, \bibinfo {author} {\bibfnamefont {N.}~\bibnamefont {Goldenfeld}},\
  and\ \bibinfo {author} {\bibfnamefont {J.}~\bibnamefont {Dantzig}},\
  }\href@noop {} {\bibfield  {journal} {\bibinfo  {journal} {Phys. rev. E 79,
  035701R}\ } (\bibinfo {year} {2009})}\BibitemShut {NoStop}%
\bibitem [{\citenamefont {Toner}\ and\ \citenamefont {Tu}(1995)}]{Toner1995}%
  \BibitemOpen
  \bibfield  {author} {\bibinfo {author} {\bibfnamefont {J.}~\bibnamefont
  {Toner}}\ and\ \bibinfo {author} {\bibfnamefont {Y.}~\bibnamefont {Tu}},\
  }\href@noop {} {\bibfield  {journal} {\bibinfo  {journal} {Phys. Rev. Lett.
  75, 4326}\ } (\bibinfo {year} {1995})}\BibitemShut {NoStop}%
\bibitem [{\citenamefont {Liu}\ \emph {et~al.}(2013)\citenamefont {Liu},
  \citenamefont {Doelman}, \citenamefont {Rottsch{\"a}fer}, \citenamefont
  {Jager}, \citenamefont {Herman}, \citenamefont {Rietkerk},\ and\
  \citenamefont {van~de Koppel}}]{Liu2013}%
  \BibitemOpen
  \bibfield  {author} {\bibinfo {author} {\bibfnamefont {Q.}~\bibnamefont
  {Liu}}, \bibinfo {author} {\bibfnamefont {A.}~\bibnamefont {Doelman}},
  \bibinfo {author} {\bibfnamefont {V.}~\bibnamefont {Rottsch{\"a}fer}},
  \bibinfo {author} {\bibfnamefont {M.}~\bibnamefont {Jager}}, \bibinfo
  {author} {\bibfnamefont {P.~M.~J.}\ \bibnamefont {Herman}}, \bibinfo {author}
  {\bibfnamefont {M.}~\bibnamefont {Rietkerk}},\ and\ \bibinfo {author}
  {\bibfnamefont {J.}~\bibnamefont {van~de Koppel}},\ }\href@noop {} {\bibfield
   {journal} {\bibinfo  {journal} {PNAS 110 (29) 11905-11910}\ } (\bibinfo
  {year} {2013})}\BibitemShut {NoStop}%
\bibitem [{\citenamefont {Theraulaz}\ \emph {et~al.}(2002)\citenamefont
  {Theraulaz}, \citenamefont {Bonabeau}, \citenamefont {Nicolis}, \citenamefont
  {Sol{\' e}}, \citenamefont {Fourcassi{\' e}}, \citenamefont {Blanco},
  \citenamefont {Fournier}, \citenamefont {Joly}, \citenamefont {Fern{\'
  a}ndez}, \citenamefont {Grimal}, \citenamefont {Dalle},\ and\ \citenamefont
  {Deneubourg}}]{Theraulaz2002}%
  \BibitemOpen
  \bibfield  {author} {\bibinfo {author} {\bibfnamefont {G.}~\bibnamefont
  {Theraulaz}}, \bibinfo {author} {\bibfnamefont {E.}~\bibnamefont {Bonabeau}},
  \bibinfo {author} {\bibfnamefont {S.~C.}\ \bibnamefont {Nicolis}}, \bibinfo
  {author} {\bibfnamefont {R.~V.}\ \bibnamefont {Sol{\' e}}}, \bibinfo {author}
  {\bibfnamefont {V.}~\bibnamefont {Fourcassi{\' e}}}, \bibinfo {author}
  {\bibfnamefont {S.}~\bibnamefont {Blanco}}, \bibinfo {author} {\bibfnamefont
  {R.}~\bibnamefont {Fournier}}, \bibinfo {author} {\bibfnamefont
  {J.}~\bibnamefont {Joly}}, \bibinfo {author} {\bibfnamefont {P.}~\bibnamefont
  {Fern{\' a}ndez}}, \bibinfo {author} {\bibfnamefont {A.}~\bibnamefont
  {Grimal}}, \bibinfo {author} {\bibfnamefont {P.}~\bibnamefont {Dalle}},\ and\
  \bibinfo {author} {\bibfnamefont {J.}~\bibnamefont {Deneubourg}},\
  }\href@noop {} {\bibfield  {journal} {\bibinfo  {journal} {PNAS 99 (15)
  9645-9649}\ } (\bibinfo {year} {2002})}\BibitemShut {NoStop}%
\bibitem [{\citenamefont {Liu}\ \emph {et~al.}(2016)\citenamefont {Liu},
  \citenamefont {Rietkerk}, \citenamefont {Herman}, \citenamefont {Piersma},
  \citenamefont {Fryxell},\ and\ \citenamefont {van~de Koppel}}]{Liu2016}%
  \BibitemOpen
  \bibfield  {author} {\bibinfo {author} {\bibfnamefont {Q.}~\bibnamefont
  {Liu}}, \bibinfo {author} {\bibfnamefont {M.}~\bibnamefont {Rietkerk}},
  \bibinfo {author} {\bibfnamefont {P.~M.~J.}\ \bibnamefont {Herman}}, \bibinfo
  {author} {\bibfnamefont {T.}~\bibnamefont {Piersma}}, \bibinfo {author}
  {\bibfnamefont {J.~M.}\ \bibnamefont {Fryxell}},\ and\ \bibinfo {author}
  {\bibfnamefont {J.}~\bibnamefont {van~de Koppel}},\ }\href@noop {} {\bibfield
   {journal} {\bibinfo  {journal} {Physics of Life Reviews 19 107–121}\ }
  (\bibinfo {year} {2016})}\BibitemShut {NoStop}%
\bibitem [{\citenamefont {Schaller}\ \emph {et~al.}(2010)\citenamefont
  {Schaller}, \citenamefont {Weber}, \citenamefont {Semmrich}, \citenamefont
  {Frey},\ and\ \citenamefont {Bausch}}]{Schaller2010}%
  \BibitemOpen
  \bibfield  {author} {\bibinfo {author} {\bibfnamefont {V.}~\bibnamefont
  {Schaller}}, \bibinfo {author} {\bibfnamefont {C.}~\bibnamefont {Weber}},
  \bibinfo {author} {\bibfnamefont {C.}~\bibnamefont {Semmrich}}, \bibinfo
  {author} {\bibfnamefont {E.}~\bibnamefont {Frey}},\ and\ \bibinfo {author}
  {\bibfnamefont {A.}~\bibnamefont {Bausch}},\ }\href@noop {} {\bibfield
  {journal} {\bibinfo  {journal} {Nature volume 467, pages 73–77}\ }
  (\bibinfo {year} {2010})}\BibitemShut {NoStop}%
\bibitem [{\citenamefont {Solon}\ and\ \citenamefont
  {Tailleur}(2013)}]{Solon2013}%
  \BibitemOpen
  \bibfield  {author} {\bibinfo {author} {\bibfnamefont {A.}~\bibnamefont
  {Solon}}\ and\ \bibinfo {author} {\bibfnamefont {J.}~\bibnamefont
  {Tailleur}},\ }\href@noop {} {\bibfield  {journal} {\bibinfo  {journal}
  {Phys. Rev. Lett. 111, 078101}\ } (\bibinfo {year} {2013})}\BibitemShut
  {NoStop}%
\bibitem [{\citenamefont {Schnitzer}(1993)}]{Schnitzer1993}%
  \BibitemOpen
  \bibfield  {author} {\bibinfo {author} {\bibfnamefont {M.~J.}\ \bibnamefont
  {Schnitzer}},\ }\href@noop {} {\bibfield  {journal} {\bibinfo  {journal}
  {PRE}\ } (\bibinfo {year} {1993})}\BibitemShut {NoStop}%
\bibitem [{\citenamefont {Bonner}\ \emph {et~al.}(1950)\citenamefont {Bonner},
  \citenamefont {Clarke(Jr)}, \citenamefont {Neely(Jr)},\ and\ \citenamefont
  {Slifkin}}]{Bonner1950}%
  \BibitemOpen
  \bibfield  {author} {\bibinfo {author} {\bibfnamefont {J.~T.}\ \bibnamefont
  {Bonner}}, \bibinfo {author} {\bibfnamefont {W.~W.}\ \bibnamefont
  {Clarke(Jr)}}, \bibinfo {author} {\bibfnamefont {C.~L.}\ \bibnamefont
  {Neely(Jr)}},\ and\ \bibinfo {author} {\bibfnamefont {M.~K.}\ \bibnamefont
  {Slifkin}},\ }\href@noop {} {\bibfield  {journal} {\bibinfo  {journal}
  {Journal of Cellular and Comparative Physiology}\ } (\bibinfo {year}
  {1950})}\BibitemShut {NoStop}%
\bibitem [{\citenamefont {Fisher}\ \emph {et~al.}(1981)\citenamefont {Fisher},
  \citenamefont {Keith},\ and\ \citenamefont {Williams}}]{Fisher1981}%
  \BibitemOpen
  \bibfield  {author} {\bibinfo {author} {\bibfnamefont {P.~R.}\ \bibnamefont
  {Fisher}}, \bibinfo {author} {\bibfnamefont {E.~S.}\ \bibnamefont {Keith}},\
  and\ \bibinfo {author} {\bibfnamefont {L.}~\bibnamefont {Williams}},\
  }\href@noop {} {\bibfield  {journal} {\bibinfo  {journal} {Cell}\ } (\bibinfo
  {year} {1981})}\BibitemShut {NoStop}%
\bibitem [{\citenamefont {Ragatz}\ \emph {et~al.}(1965)\citenamefont {Ragatz},
  \citenamefont {Jiang}, \citenamefont {Bauer},\ and\ \citenamefont
  {Gest}}]{Ragatz1965}%
  \BibitemOpen
  \bibfield  {author} {\bibinfo {author} {\bibfnamefont {L.}~\bibnamefont
  {Ragatz}}, \bibinfo {author} {\bibfnamefont {Z.}~\bibnamefont {Jiang}},
  \bibinfo {author} {\bibfnamefont {C.~E.}\ \bibnamefont {Bauer}},\ and\
  \bibinfo {author} {\bibfnamefont {H.}~\bibnamefont {Gest}},\ }\href@noop {}
  {\bibfield  {journal} {\bibinfo  {journal} {Archives of Microbiology}\ }
  (\bibinfo {year} {1965})}\BibitemShut {NoStop}%
\bibitem [{\citenamefont {Bahat}\ \emph {et~al.}(2003)\citenamefont {Bahat},
  \citenamefont {Tur-Kaspa}, \citenamefont {Gakamsky}, \citenamefont
  {Giojalas}, \citenamefont {Breitbart},\ and\ \citenamefont
  {Eisenbach}}]{Bahat2003}%
  \BibitemOpen
  \bibfield  {author} {\bibinfo {author} {\bibfnamefont {A.}~\bibnamefont
  {Bahat}}, \bibinfo {author} {\bibfnamefont {I.}~\bibnamefont {Tur-Kaspa}},
  \bibinfo {author} {\bibfnamefont {A.}~\bibnamefont {Gakamsky}}, \bibinfo
  {author} {\bibfnamefont {L.~C.}\ \bibnamefont {Giojalas}}, \bibinfo {author}
  {\bibfnamefont {H.}~\bibnamefont {Breitbart}},\ and\ \bibinfo {author}
  {\bibfnamefont {M.}~\bibnamefont {Eisenbach}},\ }\href@noop {} {\bibfield
  {journal} {\bibinfo  {journal} {Nature Medicine}\ } (\bibinfo {year}
  {2003})}\BibitemShut {NoStop}%
\bibitem [{\citenamefont {Hedgecock}\ and\ \citenamefont
  {Russell}(1975)}]{Hedgecock1975}%
  \BibitemOpen
  \bibfield  {author} {\bibinfo {author} {\bibfnamefont {E.~M.}\ \bibnamefont
  {Hedgecock}}\ and\ \bibinfo {author} {\bibfnamefont {R.~L.}\ \bibnamefont
  {Russell}},\ }\href@noop {} {\bibfield  {journal} {\bibinfo  {journal}
  {PNAS}\ } (\bibinfo {year} {1975})}\BibitemShut {NoStop}%
\bibitem [{\citenamefont {Wurtsbaugh}\ and\ \citenamefont
  {Neverman}(1988)}]{Wurtsbaugh1988}%
  \BibitemOpen
  \bibfield  {author} {\bibinfo {author} {\bibfnamefont {W.~A.}\ \bibnamefont
  {Wurtsbaugh}}\ and\ \bibinfo {author} {\bibfnamefont {D.}~\bibnamefont
  {Neverman}},\ }\href@noop {} {\bibfield  {journal} {\bibinfo  {journal}
  {Nature}\ } (\bibinfo {year} {1988})}\BibitemShut {NoStop}%
\bibitem [{\citenamefont {Brockerhoff}\ \emph {et~al.}(1995)\citenamefont
  {Brockerhoff}, \citenamefont {Hurley}, \citenamefont {Janssen-Bienhold},
  \citenamefont {Neuhauss}, \citenamefont {Driever},\ and\ \citenamefont
  {Dowling}}]{Brockerhoff1995}%
  \BibitemOpen
  \bibfield  {author} {\bibinfo {author} {\bibfnamefont {S.~E.}\ \bibnamefont
  {Brockerhoff}}, \bibinfo {author} {\bibfnamefont {J.~B.}\ \bibnamefont
  {Hurley}}, \bibinfo {author} {\bibfnamefont {U.}~\bibnamefont
  {Janssen-Bienhold}}, \bibinfo {author} {\bibfnamefont {S.~C.}\ \bibnamefont
  {Neuhauss}}, \bibinfo {author} {\bibfnamefont {W.}~\bibnamefont {Driever}},\
  and\ \bibinfo {author} {\bibfnamefont {J.~E.}\ \bibnamefont {Dowling}},\
  }\href@noop {} {\bibfield  {journal} {\bibinfo  {journal} {PNAS}\ } (\bibinfo
  {year} {1995})}\BibitemShut {NoStop}%
\bibitem [{\citenamefont {Farrell}\ \emph {et~al.}(2012)\citenamefont
  {Farrell}, \citenamefont {Marchetti}, \citenamefont {Marenduzzo},\ and\
  \citenamefont {Tailleur}}]{Farrell2012}%
  \BibitemOpen
  \bibfield  {author} {\bibinfo {author} {\bibfnamefont {F.~D.~C.}\
  \bibnamefont {Farrell}}, \bibinfo {author} {\bibfnamefont {M.~C.}\
  \bibnamefont {Marchetti}}, \bibinfo {author} {\bibfnamefont {D.}~\bibnamefont
  {Marenduzzo}},\ and\ \bibinfo {author} {\bibfnamefont {J.}~\bibnamefont
  {Tailleur}},\ }\href@noop {} {\bibfield  {journal} {\bibinfo  {journal}
  {Phys. Rev. Lett. 108, 248101}\ } (\bibinfo {year} {2012})}\BibitemShut
  {NoStop}%
\end{thebibliography}%

\end{document}